\documentclass[conference]{IEEEtran}
\IEEEoverridecommandlockouts
\usepackage{cite}
\usepackage{amsmath,amssymb,amsfonts}
\usepackage{algorithmic}
\usepackage{graphicx}
\usepackage{textcomp}
\usepackage{xcolor}
\usepackage{balance}

\usepackage{xcolor}
\definecolor{linkblue}{RGB}{0, 86, 152}
\usepackage{hyperref}
\hypersetup{
    colorlinks = true,
    linkcolor = linkblue,
    citecolor = linkblue,
    urlcolor = linkblue,
    allbordercolors={white}
}
\usepackage{nameref}
\usepackage[T1]{fontenc}
\usepackage{soul}
\usepackage{colortbl}
\usepackage{tabularx}
\usepackage{booktabs}
\usepackage{orcidlink}

\usepackage{capt-of,etoolbox}

\newcommand{
\insertfig}{
\includegraphics[width=\linewidth]{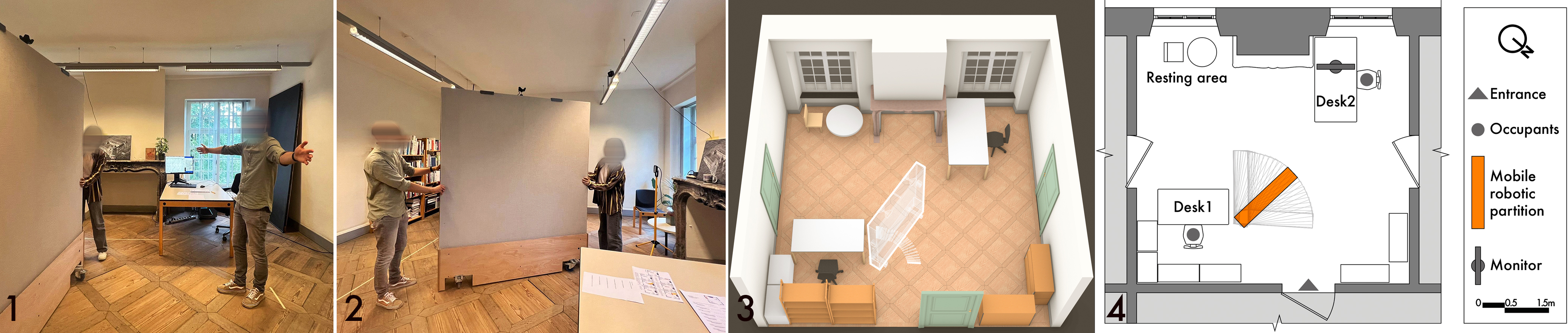}
\captionof{figure}{
During improvisation sessions, 15 experts designed gestures for a mobile robotic partition by first bodystorming how to convey an intent (1), then manually manoeuvring the partition with a researcher to perform the accordant gesture (2), before reviewing the 3D visualisation of the gesture as recorded via a real-time motion-capture system (3), and comparing all designed gestures in a questionnaire while viewing their 2D animations (4) side-by-side.
}
\label{fig:teaser}
}

\makeatletter
\apptocmd{\@maketitle}{\centering\setcounter{figure}{0}\insertfig\vspace{-0.3cm}}{}{}

\newcommand{\linebreakand}{%
  \end{@IEEEauthorhalign}
  \hfill\mbox{}\par
  \mbox{}\hfill\begin{@IEEEauthorhalign}
}

\makeatother

\def\BibTeX{{\rm B\kern-.05em{\sc i\kern-.025em b}\kern-.08em
    T\kern-.1667em\lower.7ex\hbox{E}\kern-.125emX}}
    
\begin{document}

\title{Eliciting Understandable Architectonic Gestures for Robotic Furniture through Co-Design Improvisation}

\author{
\IEEEauthorblockN{Alex Binh Vinh Duc Nguyen\orcidlink{0000-0001-5026-474X}}
\IEEEauthorblockA{
\textit{KU Leuven}\\
Leuven, Belgium \\
\href{mailto:alex.nguyen@kuleuven.be}{\nolinkurl{alex.nguyen@kuleuven.be}}}
\and
\IEEEauthorblockN{Jan Leusmann\orcidlink{0000-0001-9700-5868}}
\IEEEauthorblockA{
\textit{LMU Munich}\\
Munich, Germany \\
\href{mailto:jan.leusmann@ifi.lmu.de}{\nolinkurl{jan.leusmann@ifi.lmu.de}}}
\and
\IEEEauthorblockN{Sven Mayer\orcidlink{0000-0001-5462-8782}}
\IEEEauthorblockA{
\textit{LMU Munich}\\
Munich, Germany \\
\href{mailto:info@sven-mayer.com}{\nolinkurl{info@sven-mayer.com}}}
\and
\IEEEauthorblockN{Andrew Vande Moere\orcidlink{0000-0002-0085-4941}}
\IEEEauthorblockA{
\textit{KU Leuven}\\
Leuven, Belgium \\
\href{mailto:andrew.vandemoere@kuleuven.be}{\nolinkurl{andrew.vandemoere@kuleuven.be}}}
}

\maketitle

\begin{abstract}
The vision of adaptive architecture proposes that robotic technologies could enable interior spaces to physically transform in a bidirectional interaction with occupants. Yet, it is still unknown how this interaction could unfold in an understandable way. Inspired by HRI studies where robotic furniture gestured intents to occupants by deliberately positioning or moving in space, we hypothesise that adaptive architecture could also convey intents through gestures performed by a mobile robotic partition. To explore this design space, we invited 15 multidisciplinary experts to join co-design improvisation sessions, where they manually manoeuvred a deactivated robotic partition to design gestures conveying six architectural intents that varied in purpose and urgency. Using a gesture elicitation method alongside motion-tracking data, a Laban-based questionnaire, and thematic analysis, we identified 20 unique gestural strategies. Through categorisation, we introduced architectonic gestures as a novel strategy for robotic furniture to convey intent by indexically leveraging its spatial impact, complementing the established deictic and emblematic gestures. Our study thus represents an exploratory step toward making the autonomous gestures of adaptive architecture more legible. By understanding how robotic gestures are interpreted based not only on their motion but also on their spatial impact, we contribute to bridging HRI with Human-Building Interaction research.
\end{abstract}


\vspace{-5pt}

\section{Introduction}
The vision of \textit{adaptive architecture} proposed that the integration of robotic technologies could enable architectural spaces to physically transform in response to occupant needs~\cite{Bier2014, Fox2016}. By doing so, architectural spaces could potentially engage in bidirectional interactions with occupants~\cite{Oosterhuis2008, Schnadelbach2016, Wang2022}, offering compelling experiences~\cite{Achten2013, Meagher2015}, or even `nudging' their behaviours toward ergonomic or hedonic benefits~\cite{Green2016}. Despite these visions, the implementation of adaptive architecture in everyday spaces remains limited~\cite{Maia2015}. Beyond obvious technological challenges of sensing~\cite{Schnadelbach2019} or operating elements on an architectural scale~\cite{Urquhart2019}, one obvious inhibitor is how occupants already struggle to comprehend the autonomous intents of smart building systems. Even popular systems like sunshades, HVAC, or lighting frequently led to discomfort~\cite{Navarro2020}, a perceived lack of control~\cite{Obrien2014}, or social tension~\cite{Lashina2019} when occupants misunderstood the intents behind their actuation~\cite{Fabi2017}. It is thus still unknown how adaptive architecture and human-building interaction (HBI) systems~\cite{Alavi2016, AlaviHBI2019, Wiberg2020} in general, can communicate intents to occupants through how they unfold rather than supplementary interfaces that might be distracting or require additional effort to understand.

Recent advancements in \textit{robotic furniture} demonstrated how mobile robots integrated with everyday furniture like trashcans~\cite{Brown2014}, footstools~\cite{Sirkin2015}, chairs~\cite{LC2024}, desks~\cite{Takashima2015}, or sofas~\cite{Spadafora2016} could \textit{gesture} understandable intents toward occupants by deliberately positioning or moving in space. Inspired by these examples, we hypothesise that adaptive architecture could similarly incorporate such gestural techniques. More practically, we hypothesise that adaptive architecture can be facilitated by an autonomous mobile robotic partition able to gesture \textit{architectural intents} to co-located occupants, such as informing them of its repositioning purpose to support their activity, or nudging them to reposition themselves to a more optimal area for their activity. We chose a partition as an embodiment of adaptive architecture because partitions are among the most impactful architectural elements, shaping the perception of occupants on view, privacy, proportion, lighting, ventilation, among others~\cite{Nguyen2021, Nguyen2022, Onishi2022}. The mobility of a robotic partition would maximise its flexibility in performing gestures, unlike (semi-)fixed elements like doors, windows, or ceilings. While prior studies on robotic partitions have involved occupants~\cite{Nguyen2024}, designers~\cite{Wang2022}, or architects~\cite{Onishi2022} to design gestures as extensions of occupants~\cite{Nguyen2024, Onishi2022} or autonomous agents~\cite{Lee2013, Wang2020}, the gestural design space to convey more abstract architectural intents remains largely unexplored. Thus, our study addresses the knowledge gap: while smaller furniture robots have successfully engaged occupants through gestural performances, it remains unclear how they would transfer to adaptive architecture with larger, space-defining robots.

Our exploratory study aimed to capture a preliminary design space for how a robotic partition could gesture architectural intents toward occupants in understandable ways. Drawing inspiration from an HCI gesture elicitation method~\cite{Wobbrock2009}, where intuitive gestures for interacting with technology were rapidly generated by involving end-users~\cite{Ruiz2011, VillarrealNarvaez2020, VillarrealNarvaez2024}, we invited 15 experts to join design improvisation sessions~\cite{Hoffman2014, Sirkin2014}, during which they simulated robotic partition gestures by manually manoeuvring it. By involving experts in architecture, motion design, and robotics, we aimed to capture diverse perspectives on robotic gestures, as these perspectives often contradict each other~\cite{Johansen2019, Ribeiro2020}. The experts were tasked with designing gestures for two hypothetical occupants sharing an office space to convey six distinct intents, varying in two established HRI dimensions of \textit{purpose} and \textit{urgency}. We selected a shared office as it forms an ideal setting, hosting multiple activities that, however, may require contrasting spatial qualities.

By analysing the manually designed gestures using motion-tracking data, a questionnaire grounded in the Laban efforts~\cite{Laban1975}, and a thematic analysis of expert design reasoning, we identified 20 gestural strategies. Categorising these strategies based on their semiotic~\cite{Crow2003} differences, we proposed \textit{architectonic gestures} as a novel strategy for any robotic furniture to convey intents by self-referencing its own spatial impact in an indexical way. 
We contrast this with established strategies, such as deictic gestures that iconographically mimic anthropomorphic `pointing', or emblematic gestures that rely on a learnt symbolic `gestural vocabulary' to convey intents.

This study represents our first exploratory 
step toward enabling autonomous adaptive architecture to behave in more 
legible ways~\cite{Dragan2013}, which is expected to enhance occupant acceptance~\cite{Naneva2020} and foster longer-term human-robot trust~\cite{Edmonds2019}. Our contributions are threefold. First, we outlined a preliminary design space 
from experts across multiple disciplines on robotic partition gestures. Second, we introduced a systematic methodology for designing robotic gestures, grounded in established mixed-method 
approaches like co-design with experts, design improvisation, gesture elicitation, and Laban effort analysis. Third, we proposed architectonic gestures as a novel approach for robotic furniture to convey intents indexically. By recognising how robotic gestures are interpreted based on both their motion characteristics and spatially contextualised meaning, our study contributes to bridging HRI with HBI research in general, and architectural design in particular.

\section{Related works}
Our study is grounded on how previous furniture robots deployed gestures to interact with occupants, and the various methods through which robotic gestures were designed.

\subsection{Mobile robotic furniture gestures}
Occupants understood the gestures of a mobile furniture robot when these gestures were intuitively linked to the function of the furniture itself.  For example, a robotic door opening at varying speeds and trajectories conveyed different levels of approachability, mimicking how a person might open or close it~\cite{Ju2009}. A chair moving back and forth invited occupants to sit, similar to how one might adjust a chair~\cite{Agnihotri2019}. Likewise, a trashcan approaching occupants encouraged trash disposal, akin to a city worker offering a garbage bag~\cite{Brown2024}.

Conversely, occupants may require learning to understand more anthropomorphic gestures from furniture robots. A footstool performing attention-seeking gestures like lifting, wiggling, or bumping prompted some to rest their feet, while others viewed it as a living organism and refrained from doing so~\cite{Sirkin2015}. A toy box designed to encourage children to tidy up by repetitively pointing at toy elicited playful behaviours, instead, staying stationary and wiggling only when a toy was placed inside successfully promoted tidying~\cite{Fink2014}. When carefully directed, anthropomorphic gestures can help furniture robots, like a sofa, convey compelling `personalities'~\cite{Spadafora2016}.

In shared spaces, co-located occupants intuitively understood gestures of furniture robots when they navigated between personal proxemic zones~\cite{Hall1969}. A pair of tables encouraged two occupants to approach each other by gradually merging their personal proxemic zones~\cite{Takashima2015}. A pair of bar stools rotated two occupants to face one another, subtly promoting social interaction~\cite{Sadka2022}. In a shared office, a robotic partition gestured workers to modify disturbance-causing behaviours by moving towards either their own desks or affected colleagues~\cite{Nguyen2024}.

\subsection{Designing robotic gestures}
Since non-humanoid robots cannot directly replicate~\cite{VandePerre2015} or mimic human gestures~\cite{Suguitan2024}, their gestures are typically developed through \textbf{co-design with experts}~\cite{Abe2022}. Subtle gestures of a minimalist robot were co-designed with animators, puppeteers, choreographers, and comic artists to convey user-recognisable intents~\cite{AndersonBashan2018}. With a clown therapist and an animator, humorous gestures for a lamp-like robot~\cite{Press2022} were co-designed to reduce social awkwardness among strangers~\cite{Press2023}. To align with the storyline-based approach of animators~\cite{Ribeiro2020}, co-design method often required an iterative process~\cite{Hoffman2014} that fine-tuned both the morphology and gestures of a robot until understandable intents were achieved~\cite{Hoffman2015}.

Pioneering studies of robotic furniture have outlined a research-through-design approach for generating gestures, known as \textbf{design improvisation}~\cite{Hoffman2014, Sirkin2014}, which has been applied for robotic chairs~\cite{LC2024}, trashcans~\cite{Brown2024}, footstools~\cite{Sirkin2015}, or sofas~\cite{Spadafora2016}. This approach involves first storyboarding interaction scenarios, followed by prototyping lower-fidelity Wizard-of-Oz robots. These prototypes are then used in gestural improvisation sessions with experts, which are video-recorded to simulate ecologically valid contexts. These videos are then virtually crowd-sourced for broad feedback, guiding gesture refinement for final validation in user studies~\cite{Sirkin2014}.

To identify the most intuitively understandable robotic gestures, previous HRI studies~\cite{Leusmann2024Approach} have proposed that the method of \textbf{gesture elicitation}~\cite{Wobbrock2009} is highly suitable. Widely used in HCI~\cite{Ruiz2011, VillarrealNarvaez2020, VillarrealNarvaez2024}, this method involves lay-participants generating gestures corresponding to a set of referents, which are originally features of a computing system. Researchers then merge similar gestures, calculate their frequency, and select the most frequent gesture for each referent. For robotic furniture, these referents could instead represent the intents that the gestures are meant to convey.

Given the diversity of gestures, HRI studies have employed the qualitative framework of \textbf{Laban efforts}~\cite{Laban1975} from choreography to systematically describe them~\cite{Venture2019}. Originally developed to record and choreograph human motion, the Laban efforts encompass four dimensions: \textit{space} (motion shape), \textit{weight} (exerted energy), \textit{time} (rhythm or tempo), and \textit{flow} (motion connectedness). This framework has been applied to design gestures for both stationary~\cite{Galindo2022, Viola2022} and mobile~\cite{Knight2014, Knight2016, Peng2020, Emir2022} robots, including furniture robots like mobile chairs~\cite{Knight2017, Fallatah2019}, demonstrating how gestures varied in Laban dimensions can convey distinct intents~\cite{Knight2017}.

\section{Methodology}
We invited 15 experts to design improvisation sessions~\cite{Hoffman2014, Sirkin2014} for robotic partition gestures. We then applied the gesture elicitation method~\cite{Wobbrock2009} to distil core gestural strategies among designed gestures and benchmarked their differences with a Laban-based questionnaire~\cite{Laban1975, Knight2014}.

\subsection{Architectural context}
Our study was conducted in a real office located on our university campus. As illustrated in \autoref{fig:teaser}, this office provided ample space (\(36m^2\)) to host two work desks and a resting area, with a single entrance and two additional doors leading to storage rooms. Two large southwest-facing windows and a high ceiling allow for abundant natural light and offer expansive views of the surroundings. The combination of traditional European architectural features with minimalist furniture and visible technical elements together creates a postmodern yet cosy atmosphere. We arranged two work desks and a resting area at three distinct corners of the room, allowing occupants in any area to enjoy the window views while leaving sufficient space for the partition to be manoeuvred in between. 

\subsection{Technical implementation}
Within the study office, we deployed a robotic partition as shown in \autoref{fig:teaser}. 
Measuring \(180x210x28cm\), it was large enough to obstruct the view from a work desk to other areas of the office, while still being able to move freely without collision. To enable experts to comfortably manoeuvre the partition, the robotic wheels were deactivated. However, the weight of the acoustic panels still introduced a realistic level of resistance onto these wheels, ensuring that the physical constraints were accounted for during gesture improvisation.

We implemented a customised motion-tracking system to capture improvised gestures. The hardware setup included two Vive\footnote{Vive: \href{https://www.vive.com/}{https://www.vive.com/}} Base Stations, each mounted at a height of \(2.5m\) in opposite corners of the office, and a Vive Tracker affixed to the top of the partition at \(1.9m\), ensuring its continuous visibility. Developed for tracking gestures of virtual reality users, this system offers precise capturing of improvised gestures with an accuracy of up to \(7mm\)~\cite{Holzwarth2021}. 
We implemented the tracking system software on a computer on one of the desks, as shown in \autoref{fig:teaser}. A \(Python\) script continuously logged the tracker position and orientation every 2 milliseconds. A \(Grassopper\)\footnote{Grasshopper 3D: \href{https://www.grasshopper3d.com/}{https://www.grasshopper3d.com/}} script streamed the data to visualise the gesture in real time. We rendered this visualisation within a 3D model of the office in \(Rhino\)\footnote{Rhino 3D: \href{https://www.rhino3d.com/}{https://www.rhino3d.com/}},
displaying the gesture as if it were autonomously performed by the partition. 
Another \(Grasshopper\) script then enabled a side-by-side comparison of multiple gestures by animating them on 2D-floor plans, as shown in \autoref{fig:teaser}. 

\begin{table}[t]
    \centering
    \scriptsize
    \caption{The six robotic partition intents that were shown to experts as referents during gesture improvisation.}
    \label{tab:intent}
    \begin{tabularx}{\linewidth}{@{}p{0.8cm}p{0.3cm}p{0.1cm}X@{}}
        \toprule
        \textbf{Purpose} & \textbf{Urgency} & \multicolumn{2}{c}{\textbf{Intent referent}} \\
        \midrule
        Informing & Non-urgent & \textit{I1} & The partition informs the occupant that it will move to cover the glare for them. \\
        Informing & Urgent & \textit{I2} & The partition informs the occupant that it will move urgently to cover the glare for the other occupant.  \\
        Nudging & Non-urgent & \textit{I3} & The partition suggests the occupant relocate to the resting area to read, but only if they want to. \\
        Nudging & Urgent & \textit{I4} & The partition suggests the occupant move outside urgently for their phone call to avoid disturbing the other occupant. \\
        \multicolumn{2}{c}{Control: Availability} & \textit{I5} & The partition is available for any occupant to use. \\
        \multicolumn{2}{c}{Control: Uncertainty} & \textit{I6} & The partition is unsure where to move to. \\
        \bottomrule
    \end{tabularx}
    \vspace{-15pt}
\end{table}

\subsection{Gesture improvisation}

\subsubsection{Intents}
We selected six intents (I1-6) as referents for the gesture improvisation, as detailed in \autoref{tab:intent}, in which four are based on two established HRI dimensions.
The first dimension, \textbf{purpose}, assessed the impact of robotic autonomy~\cite{Beer2014}, where the partition either informed the occupants of the reason for its action or `nudged' them to perform actions themselves. For \textit{informing} purposes, we selected covering glare from sunlight as the reason (I1,2). Glare was not only an existing disturbance in the selected office, but it also motivated workers to utilise a partition in a previous study~\cite{Nguyen2024}. For \textit{nudging} purposes, we aimed to encourage an occupant to relocate (I3,4), as we hypothesised this purpose might inspire experts to design diversely different gestures.
The second dimension, \textbf{urgency}, assessed how experts would modify a similar gesture to convey either urgent or non-urgent intent. To legitimise the sense of urgency, we chose \textit{urgent} intents as the partition informing or nudging an occupant to assist the other occupant (I2,4), while \textit{non-urgent} intents aimed to assist an occupant directly (I1,3).
To better benchmark the two dimensions above, we included two additional \textbf{control} intents (I5,6), not linked to purpose or urgency but representing two typical robotic functionalities: conveying availability versus uncertainty.

\subsubsection{Measures}\label{sec:questionnaire}
\begin{figure}[t]
    \centering
    \includegraphics[width=\linewidth]{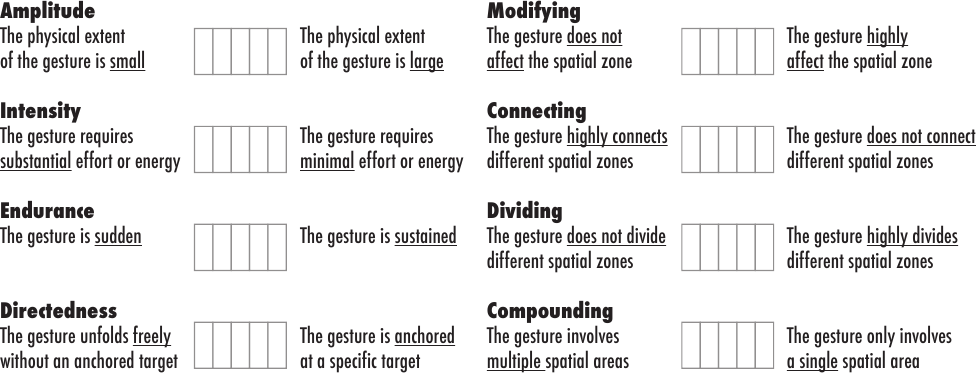}
    \caption{The questionnaire used to elicit experts in differentiating their own designed gestures, with the order of measures randomised for each expert.
    }
    \label{fig:questionnaireform}
    \vspace{-12pt}
\end{figure}

To capture how experts self-differentiate their designed gestures, we developed a questionnaire consisting of eight five-point Likert-type measures, as shown in \autoref{fig:questionnaireform}.
The first four measures were adapted from the Laban effort dimensions~\cite{Laban1975} to capture the \textbf{dynamic} characteristics of a gesture. These measures included \textit{amplitude} (adapted from Laban `space'), assessing the physical extent of the gesture; \textit{intensity} (from Laban `weight'), measuring the perceived energy exerted by the gesture; \textit{endurance} (from Laban `time'), evaluating the temporal rhythm of the gesture; and \textit{directedness} (from Laban `flow'), capturing whether the gesture unfolded freely or was anchored towards a specific point. The other four measures, based on prior studies of robotic partitions~\cite{Nguyen2021}, aimed to capture the \textbf{spatial} characteristics of a gesture. These measures assessed how far the gesture \textit{modified} the qualities of a spatial area, \textit{divided} or \textit{connected} an area to another one, and \textit{compounded} multiple spatial areas as the gesture unfolded.

\subsubsection{Recruitment}
We employed purposive sampling~\cite{Campbell2020} to recruit participating experts, starting with emails sent individually to those affiliated with our university, followed by a snowball sampling~\cite{Naderifar2017} expanding through their networks. The emails explained the study objectives, duration, location, and the researchers' contact details, with a personalised message highlighting the relevance of the recipient's expertise, and the information that a 10 Euro donation would be made to a university-affiliated charity on their behalf. Interested experts were invited to register through an online form, providing optional demographic details and selecting available time slots.

\subsubsection{Procedure}
At the beginning of a 1-hour design improvisation session, the participating expert was invited to familiarise with the office space and the robotic partition by observing a video of it autonomously moving in the same space.
After the expert signed an informed consent form, the researcher explained the six intents. Each expert was tasked with designing the gestures for these intents in a randomised order. Each intent was illustrated on a printed sheet (included in the Appendix), with the initial position of the partition in an annotated floor plan, its targeted final position (for I1 and I2), and a textual description identical to \autoref{tab:intent}. For each intent, the researcher followed the instructions of the expert to together manoeuvre the partition, while the motion-tracking system recorded the gesture. If the expert struggled to ideate a gesture, they were encouraged to bodystorm~\cite{Oulasvirta2003} it themselves before translating to the partition. A movable spotlight could be used to simulate glare at any time.

After each gesture was recorded, the expert reviewed its animated 3D visualisation, from which they could choose to re-improvise if needed. Throughout the improvisation session, the researcher encouraged the expert to continuously reflect on their design choices using the think-aloud method~\cite{Charters2003}. Once all six gestures were recorded, the expert viewed their 2D animations side-by-side in the same order that they designed while completing a questionnaire for each gesture. For I1 and I2, the researcher instructed the expert to focus on evaluating the gesture conveying the informing or nudging intent, rather than the required transition of the partition from its initial to the targeted position. The session concluded with a semi-structured interview, where the expert explained their design reasoning, critiqued their designed gestures, and reflected on how lay occupants might interpret the gestures.

\subsection{Data acquisition and analysis}
The collected data include experts' questionnaire responses, audio recordings of their design reasoning during improvisations and interviews, recorded gestures from the motion-tracking system, and observation notes.
We applied the gesture elicitation method to distil designed gestures into a core set of unique \textit{gestural strategies}, each representing distinct motions aimed to convey specific meanings. Using open coding, one author grouped gestures based on their similarities in motion and design reasoning, resulting in 39 unique gestures. Through axial coding, we further condensed these gestures into 20 gestural strategies, as many gestures combined multiple strategies to convey an intent. The coding process involved frequent group discussions between two authors. For each strategy and intent combination, we calculated an occurrence score based on the frequency a strategy was designed for a given intent, in proportion to the total number of strategies for that intent, thus, higher scores indicating greater expert agreement.

To evaluate the dimensions of purpose and urgency, the \textit{questionnaire} results were categorised into five groups: gestures of informing, nudging, non-urgent, urgent, and control intents. With eight measures in the questionnaire, this categorisation resulted in 40 datasets. Following fair statistical communication guidance~\cite{Emam2011, Dragicevic2016}, we first used Cronbach's alpha to assess internal validity for five datasets in each measure. We evaluated the normality of each dataset using the Shapiro-Wilk test and Q-Q plots. As not all datasets were normally distributed, we applied the non-parametric \textit{within-subject} pairwise Wilcoxon signed-rank test to compare their differences. 
We statistically analyse the questionnaire results rather than the motion-tracking data, as the questionnaire responses more accurately captured design intentions of experts, whereas the manually-improvised gestures often lacked precision in intended speed or trajectory.
We transcribed the \textit{audio} to derive experts' design reasoning, which we then analysed following the six-phase method of reflexive thematic analysis~\cite{Braun2006}. The familiarisation phase occurred as weekly discussions between two authors before one author inductively coded the transcript into 31 codes that foregrounded main design considerations from experts. The seven identified themes were iteratively reconsidered, split, and merged via group discussions into four themes as reported in \ref{sec:reasoning}.

\section{Results}
We recruited 15 participating experts in animation (P01-3), cinematography (P04-5), choreography (P06-7), scenography (P08-9), architectural design (P10-12), and robotic development (P13-15). As each expert designed six gestures, we collected 90 total gestures (39 unique). 

\subsection{Gestural strategies}
\label{sec:patterns}

\begin{table}[t]
    \centering
    \scriptsize
    \caption{The 20 gestural strategies designed by our experts and categorised based on their types and intended intents.}
    \label{tab:patterns}
    \begin{tabularx}{\linewidth}{@{}p{0.15cm}Xp{0.7cm}@{}}
        \toprule
        \multicolumn{2}{c}{\textbf{Gestural strategies}} & \textbf{Intent} \\
        \hline
        \multicolumn{3}{c}{\cellcolor{white!85!black}\textit{Architectonic strategies}} \\
        GA1 & \textbf{Directional corridor}. Move to form a corridor-like path for the occupant to follow & I3, I4 \\
        GA2 & \textbf{Welcoming space}. Move to create a `welcoming space' for the occupant to follow & I3 \\
        GA3 & \textbf{Visual connection}. Move away to open up the visual connection between two occupants & I4, I6 \\
        \hline
        GA4 & \textbf{Close impact}. Move close to the occupant to show the biggest impact, then move to the final position & I1 \\
        GA5 & \textbf{Midway pause}. Move halfway to the intended position, pause to highlight benefits for the occupant, then continue & I1 \\
        GA6 & \textbf{Slow reveal}. Move slowly to the intended position to gradually reveal the benefits to the occupant & I1 \\
        GA7 & \textbf{Block and open}. Move to `close off' the occupant, then `open up' to encourage them to exit & I4 \\
        \hline
        \multicolumn{3}{c}{\cellcolor{white!85!black}\textit{Deictic strategies}} \\
        GD1 & \textbf{Diagonal direction}. Move to a diagonal position to show a direction for the occupant & \textbf{I1, I3, I4}\\
        \hline
        GD2 & \textbf{Width back-off}. Move away orthogonally along the width to `back off' from supporting the occupant & I2 \\
        GD3 & \textbf{Directional retreat}. Move away orthogonally along the length to indicate a direction for the occupant & I2 \\
        \hline
        GD4 & \textbf{Length point}. Linear oscillation along length as `pointing' towards a direction for the occupant & I3 \\
        GD5 & \textbf{Hinged point}. Hinged rotational oscillation as `pointing' towards a direction for the occupant & I3, I4 \\
        \hline
        \multicolumn{3}{c}{\cellcolor{white!85!black}\textit{Emblematic strategies}} \\  
        GE1 & \textbf{Custom position}. Move to a particular position that occupants can assign with a specific meaning & I1, I2, I5, I6 \\
        \hline
        GE2 & \textbf{Confused back-and-forth}. Continuously move between two occupants to convey uncertainty & I6 \\
        GE3 & \textbf{Confused head shake}. Symmetrical rotational oscillation, like `shaking head', to convey uncertainty & \textbf{I6} \\
        GE4 & \textbf{Confused look around}. Continuous rotation, like `looking around', to convey uncertainty & I6 \\
        \hline
        GE5 & \textbf{Length wave}. Linear oscillation along length as `waving' at the occupant & I4, I5 \\
        GE6 & \textbf{Hinged wave}. Hinged rotational oscillation as `waving' at the occupant & I3, \textbf{I5} \\
        GE7 & \textbf{Symmetrical wave}. Symmetrical rotational oscillation as `waving' at the occupant & \textbf{I2}, I4, I5 \\
        GE8 & \textbf{Width step-up}. Linear oscillation along width as `stepping up' for the occupant & \textbf{I5} \\
        \bottomrule
    \end{tabularx}
    \vspace{-10pt}
\end{table}

Through the gestural coding process, we distilled 20 core gestural strategies as in \autoref{tab:patterns}. We further categorised these into three types: seven \textit{architectonic strategies} that relied entirely on the spatial impact of the partition, either through its position (e.g., GA1-3) or spatial transformation unfolding through its motion (e.g., GA4-7), to convey intents; five \textit{deictic strategies} that directed occupant attention to spatial areas via anthropomorphic `pointing' gestures, such as through a diagonal position (GD1), transition (GD2-3), or oscillatory motion (GD4-5); and eight \textit{emblematic strategies}, which employed symbolic motions independent of spatial context, either self-defined by the experts (GE1) or translated from emblematic human gestures~\cite{Matsumoto2013}, such as `shaking head' (GE3), `looking around' (GE4), or `waving' (GE5-7).


\begin{figure*}[ht]
    \centering
    \includegraphics[width=\linewidth]{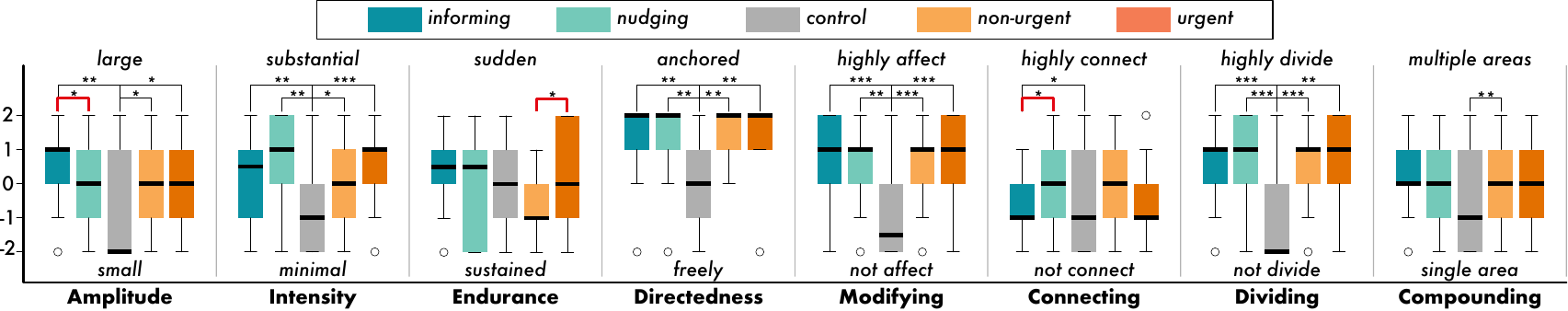}
    \caption{The questionnaire results of 15 experts evaluating their 90 designed gestures across eight measures. The gestures were categorised into five groups based on the dimensions of purpose (informing versus nudging) and urgency (non-urgent versus urgent), along with control gestures. 
    }
    \label{fig:questionnaire}
    \vspace{-10px}
\end{figure*}

As shown in \autoref{tab:patterns}, our experts employed both architectonic and deictic strategies for I1, because this intent required directional or spatial cues, such as the partition gesturing a direction toward the window (GD1) or demonstrating its glare-blocking impact (GA4–6). 
For I2, experts primarily employed emblematic and deictic strategies, as they believed the partition should perform oscillatory gestures to capture attention (GE7), but also indicate a direction toward the other occupant it would move to support (GD2–3). 
All three strategies were used for I3 and I4, as while some experts felt that emblematic `waving' were sufficient to convey nudging intents (GE5–7), others preferred clearly gesturing where the occupant should relocate through architectonic (GA1-2) or deictic (GD4-5) strategies.
For I5 and I6, emblematic strategy dominated, as oscillatory, stationary gestures were thought to communicate uncertainty (GE2–4) or availability (GE8) without relying on spatial or directional cues.
As a result, GD1 (diagonal direction) has the highest occurrence score for I1 (\(46.7\%\)), I3 (\(25.0\%\)), and I4 (\(40.9\%\)), as all three intents required clear directional cues. GE7 (symmetrical waving) was most commonly used for I2 (\(46.7\%\)) to inform the occupant that the partition was about to leave, while GE6 (hinged wave) and GE8 (width step-up) were equally preferred for I5 (\(26.7\%\)) to convey availability. For I6, GE3 (confused head shake) was the most selected strategy (\(40.0\%\)) to convey uncertainty.

\subsection{Questionnaire}
After confirming reliable internal consistency across five datasets in each measure using Cronbach's alpha (with the lowest \(alpha=0.74\)), the Wilcoxon signed-rank tests revealed that control gestures significantly differed from other gestures on nearly every measure, likely because they were more stationary, as evidenced in \ref{sec:patterns}. As shown in \autoref{fig:questionnaire}, our experts rated informing gestures as having significantly larger amplitudes than nudging gestures (\(p=0.03\)); while nudging gestures scored significantly higher than informing in terms of connecting one spatial area to another (\(p=0.02\)); and urgent gestures were perceived as significantly more sudden than non-urgent (\(p=0.01\)). These results indicate: (1) nudging intents tend to require gestures that stay close to the occupant while offering clear spatial cues connecting them to other areas; and (2) urgent intents tend to require quicker, more abrupt gestures.

\subsection{Design reasoning}
\label{sec:reasoning}

\subsubsection{Gestural inspiration}
Our experts reasoned that architectonic gestures felt ``\textit{natural}" (P10) to convey intents (P04, P09, P11) because occupants would ``\textit{immediately notice any spatial changes}" (P04), such as when the partition created a corridor (GA1) or a welcoming space (GA2). A blocking and opening gesture (GA7) was perceived as mimicking the ``\textit{open of a door}" (P07) to encourage an occupant to ``\textit{look towards and enter}" (P09).
Some experts designed architectonic gestures that relied on social cues, such as by letting the partition open up the visual connection between two occupants (GA3) to remind a phoning occupant of the disturbance they might be causing to the other (P06, P08, P09). To convey informing intents, experts believed architectonic gestures should leverage situational cues, as the partition should gesture only when an occupant is affected by glare (P06, P10, P15) or immediately reposition to cover glare before performing other motions (P05, P08, P13).
Given the ``\textit{orthogonal arrangement}" of the office layout (P08), some experts believed deictic gestures that position the partition diagonally (GD1) could capture occupant attention (P01) to guide their gaze (P04). A gradual linear motion along the partition's length (GD3) was interpreted as encouraging occupants to relocate, as ``\textit{people are unconsciously inclined to follow a directional motion}" (P06).

Other experts believed mimicking human emblematic gestures would make the partition feel ``\textit{less obtrusive}" (P08) and more ``\textit{like a soft, living element}" (P02). They chose subtle oscillatory gestures (GE5, GE8) that mimicked ``\textit{breathing, like being alive}" (P06) to indicate availability (P07, P09). Some referenced sport-related motions like the split-jumps to block a ball strike in tennis (P07, P15) while designing linear oscillatory gestures (GE5) to convey glare-covering intents. Rotational oscillatory gestures (GE2-14) were interpreted as ``\textit{confusion}" (P01), ``\textit{like a person shaking their head}" (P10); or as ``\textit{waving}" (P11) to attract attention (P14). Experts also typically assigned the partition with a personality, imagining it as a ``\textit{nice}" (P12), ``\textit{calm}" (P12), ``\textit{friendly}" (P02) robot, which however should be ``\textit{very self-aware, working to build a relationship with the occupants}'" (P01).

\begin{figure*}[ht]
    \centering
    \includegraphics[width=\linewidth]{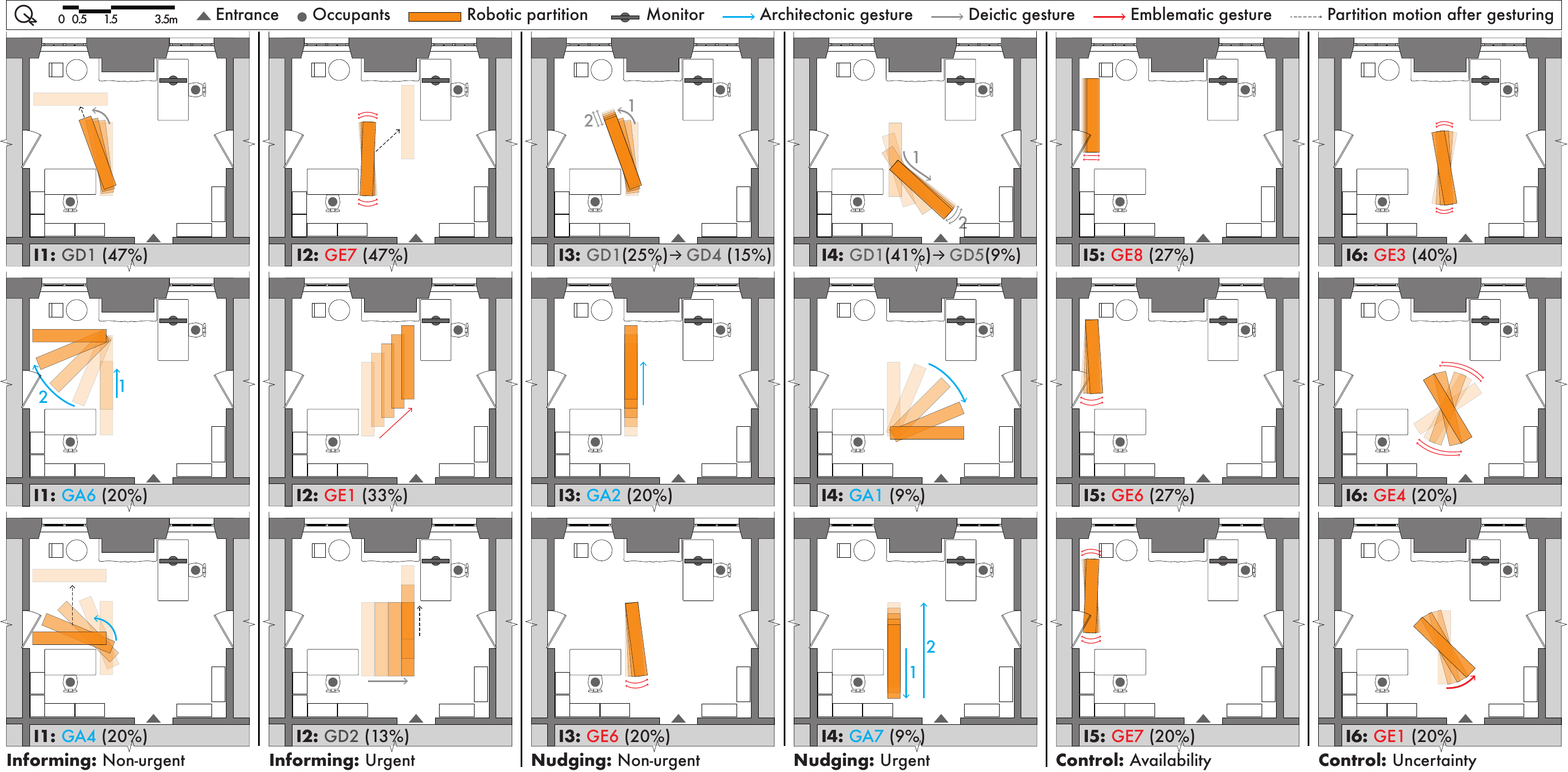}
    \caption{The three gestural strategies that were most frequently designed for each of the six intents (see \autoref{tab:intent}), together with their occurrence scores (\%), showing how `nudging' intents often required gestural sequences that combined multiple gestural strategies (see \autoref{tab:patterns}) more than other intents.}
    \label{fig:gestures}
    \vspace{-10px}
\end{figure*}

\subsubsection{Gestural interactivity}
To convey complex intents, experts designed `interaction sequences' that combined multiple gestural strategies, as shown in \autoref{fig:gestures}, which would unfold gradually based on the response of the occupant. A nudging sequence (e.g. P08, P09, P15) often began with the partition using an emblematic gesture to attract attention. It would then indicate a direction with a deictic or architectonic gesture, wait for the occupant to follow, and, if they chose not to relocate, return to its original position after some time. For informing intents, experts believed the partition should first quickly cover the glare, then follow the changing angles of sunlight to continuously protect the occupant (P06, P13). To convey availability without disturbing occupants, 11 experts suggested that the partition should perform subtle emblematic gestures intermittently, such as by oscillating every fifteen minutes (P07), half an hour (P10), at the start of the day (P09) or only when being looked at (P11). The gesture should cease if the partition remained unused after being noticed (P01).

To convey urgency, experts proposed that the partition could employ faster speeds, larger trajectories, or more explicit spatial directionality.
Ten experts believed that ``\textit{higher speed intuitively signals urgency}" (P13) because people are ``\textit{sensitive to changes in speed}" (P05). Five experts preferred using larger trajectories because bigger motions would better capture attention (P05, P10) and provide stronger motivation: ``\textit{I would feel like I need to move to match the partition's effort}" (P15). 
Three experts (P12, P13, P14) explained that by blocking the access of a phoning occupant to the rest of the office while directing them towards the exit, the partition would clearly communicate the urgency to take the call outside.

\subsubsection{Gestural understandability}
To improve gesture understandability, 13 experts developed consistent `gestural vocabularies', often assigning emblematic gestures to specific intents. When asked if occupants would appreciate this approach, experts likened it to learning ``\textit{how to use a new robot}" (P13), ``\textit{the culture of a new country}" (P10), or ``\textit{the dynamics of a new office}" (P01). However, they reflected that unfamiliar occupants might face a ``\textit{learning curve}" (P04) in ``\textit{gasping the language of the [partition]}" (P05), potentially requiring ``\textit{a training session}" (P13) or ``\textit{a handbook}" (P12). Five experts emphasised that anthropomorphic-inspired emblematic gestures like `shaking head' or `waving' might be interpreted differently depending on cultures (P10) or communities (P07).

To ensure that occupants would not interpret gestures intended for another as for them, five experts suggested that gestures should occur closer to the intended occupant (P10), always orient the larger side of the partition towards them to mimic how people face each other during conversations (P13), and involve smaller trajectories (P14, P15). To prevent occupants from mistaking gestures as robotic errors, three experts emphasised that they should be executed ``\textit{decisively}" (P10) with ``\textit{smooth}" (P01) or ``\textit{efficient}" (P11) trajectories.

\subsubsection{Gestural requirement}
All experts acknowledged that any gesture could inadvertently distract occupants, particularly if it lasts for too long (\(n=4\)), has a higher speed (\(n=3\)) or a larger trajectory (\(n=8\)), comes too close to an occupant (\(n=4\)), obstructs their view (\(n=2\)), or blocks their access (\(n=2\)). They thus proposed that a gesture should only occur when necessary, such as when its benefit outweighs the distraction it might cause (P01, P11), or when it is useful enough to justify disrupting work activities (P02, P09, P11).

Four experts noted that any gesture should prioritise office safety by ensuring that it does not block exits (P07, P15) or slows down when moving through circulation areas (P14). However, they reflected that these safety measures might hinder understandability. Ten experts highlighted that some gestures might only be feasible in the given office, where there is sufficient space and fewer occupants. While architectonic gestures may require an open space for manoeuvre (P01, P07, P04, P05, P13), deictic and emblematic gestures need to be visible to all occupants to capture attention (P05, P12, P13).

\section{Discussion}

\subsection{Eliciting robotic gestures through co-design improvisation}
Our findings demonstrate how our systematic methodology enabled the identification and benchmarking of diverse gestural strategies. Through \textit{co-design with experts}~\cite{Abe2022} in architecture, motion design, and robotics, we captured multifaceted insights on robotic partition gestures, including not only their spatial impact, the temporal narrative of how they should gradually unfold, but also their compatibility with particular robotic kinematics. Through \textit{design improvisation}~\cite{Hoffman2014, Sirkin2014}, experts were able to bodystorm, evaluate, and refine their gestures while providing real-time reasoning to support later thematic analysis. The architectural intents, acting as referents of the \textit{gesture elicitation} method~\cite{Wobbrock2009}, familiarised experts with more abstract purpose and urgency, while the questionnaire inspired by the \textit{Laban efforts}~\cite{Laban1975} prompted them to evaluate gestures on both dynamic and spatial aspects.

We reflect that the success of this methodology, however, depended on several factors. First, the architectural context for design improvisation needed to be carefully setup to ensure ecological validity, providing consistent, precise spatial cues for designing gestures. In our study, these cues included subtle details like sunlight intensity, glare direction, or visibility between desks. Second, experts might require sufficient familiarity with the robotic furniture to design technically feasible gestures, which was achieved in our study through a video introduction and hands-on experience manoeuvring the inactive robotic partition. Third, the questionnaire terminologies needed to be interpretable across disciplines, as our choreography experts readily understood the Laban-based measures, while architecture experts found the spatial measures more familiar. Lastly, the personal perception of experts on the robotic furniture might have influenced their designs, as while some of our experts asserted the nudging intents as beneficial, others opposed overt robotic autonomy in workplaces, choosing gestures with less explicit autonomy. While we were able to encourage opposing experts to still design gestures for these intents, future studies could benefit from a questionnaire assessing initial trust perception~\cite{Schaefer2016} of experts on the robotic furniture to gain deeper insights into their design choices.

\subsection{Architectonic gestures for robotic furniture}
\begin{figure}[t]
    \centering
    \includegraphics[width=\linewidth]{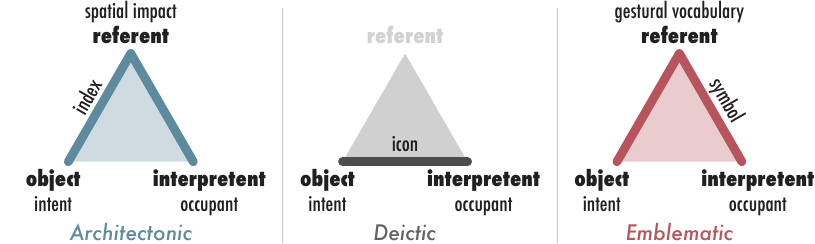}
    \caption{Our experts viewed architectonic gestures as \textit{indexically} referencing the spatial impact of the partition to convey intent, semiotically distinguishing them from \textit{iconographic} deictic gestures and \textit{symbolic} emblematic gestures.
    }
    \label{fig:semiotic}
    \vspace{-10px}
\end{figure}

Our findings identified a design space encompassing three strategies used by experts to design robotic partition gestures, each employing a distinct semiotic approach~\cite{Crow2003}. As shown in \autoref{fig:semiotic}, architectonic strategy referenced the spatial impact of the partition itself in an indexical way to convey spatially-related intents; while deictic strategy iconographically mimics anthropomorphic `pointing'; and emblematic strategy required occupants in knowing a symbolic `gestural vocabulary' to grasp their intents. While deictic and emblematic gestural strategies were well-documented in HRI~\cite{Sauppe2014, Wicke2021, Catlin2023}, our study introduced architectonic strategy as a novel approach. Our experts believed that simple repositioning of the partition could provide sufficient cues to inform or nudge occupants by spatially adjusting the directionality (GA1), visibility (GA3), or accessibility (GA7) of their surroundings; or even by leveraging surrounding contextual cues in a potentially intuitive way, such as to (dis)align the partition with the office layout or to time its gesture based on environmental changes.

The choice of our experts in designing non-repetitive, gradual, and thus less dynamic architectonic gestures was likely influenced by the office context and the partition dimensions. In a quiet, focused office, experts felt the partition should function more like a background element, leading to a resistance to more dynamic gestures. This resistance was amplified by its large dimensions, which could evoke unease or even a sense of danger if it moved .
Given that any furniture robot, whose morphology suggests furniture functionality, might be similarly perceived as a background element, we propose that their gestures could also be designed using the architectonic strategy to enable a seamless transition from static furniture to mobile robot.


\subsection{How to convey intent and urgency}
Because our experts perceived a trade-off between intrusiveness and understandability across the three gestural strategies, they recommended selecting or combining them to convey intents sufficiently. Specifically, experts reflected that emblematic gestures could captured attention but lacked spatial clarity; architectonic gestures offered spatial cues but risked `over-interpretation', with occupants attributing intent to any spatial configuration; and deictic gestures combined spatial clarity with attention-grabbing motions, but tended to amplify intrusiveness.
Some experts thus suggested combining different gestural strategies in an `interaction sequence' that would unfold gradually based on occupant response. We therefore propose future research to evaluate the hypotheses of our experts, on whether intrusiveness and understandability could be balanced by alternating or combining gestural strategies. 


To convey a sense of urgency, our experts proposed that a particular gesture could be intensified along one of the three Laban efforts~\cite{Laban1975}, such as by increasing tempo (endurance), expanding trajectory (amplitude), or exerting more explicit spatial directionality (directedness). This approach differs with previous HRI findings in two ways. First, while shorter trajectories of mobile robots were often perceived as more urgent due to their efficiency~\cite{Knight2014, Knight2016}, our experts felt the partition conveyed urgency through larger trajectories. This was because such trajectories would create more noticeable spatial changes, prompting occupants to act quickly to restore the office to its original state. Second, while Laban efforts have predominantly been used to decode robotic motion paths~\cite{Sharma2013, Knight2017, Emir2022}, our findings suggest that they can also differentiate the spatial transformations created by these motions. We propose that this shift in interpretation, from focusing only on robotic motion to also its spatial impact, may provide a bridge between HRI with HBI, potentially motivating a more seamless integration of robots into everyday environments.

\section{Conclusion and Limitations}
Through a systematic methodology that combined established mixed-method approaches - including co-design with experts, design improvisation, gesture elicitation, and Laban effort analysis - we captured a design space from multidisciplinary experts on robotic partition gestures to convey architectural intent. We identified architectonic gestural strategy as a novel approach for robotic furniture to interact with humans, leveraging its spatial impact to convey intent while enabling a seamless transition from static background elements to mobile robots. By emphasising that robotic furniture gestures should be designed not only through their motion but also spatial impact, our study contributes to bridging HRI with HBI research in general and architectural design in particular. 
However, since our study involved only experts, future research should validate how lay occupants actually perceive these gestural strategies and explore whether gestural intrusiveness and understandability could be balanced by alternating or combining them. The large scale of our partition and the quiet office setting may have favoured subtle architectonic gestures that could be ineffective for smaller robots in busier environments. Therefore, future studies should scope the applicability of architectonic gestures across different robots and contexts.

\section*{Acknowledgement}
This research is supported by the KU Leuven ID-N project IDN/22/003 ``Adaptive Architecture: the Robotic Orchestration of a Healthy Workplace" and the HORIZON-HLTH-2023-ENVHLTH-02 SONATA project. This project has received funding from the European Union's Horizon Europe Research and Innovation Actions programme under Grant Agreement no. 101137507. Views and opinions expressed are however those of the authors only and do not necessarily reflect those of the European Union or Health and Digital Executive Agency. Neither the European Union nor the granting authority can be held responsible for them.

\bibliographystyle{IEEEtranDOI}

\balance
\bibliography{main}

\end{document}


\title{Eliciting Understandable Architectonic Gestures for Robotic Furniture through Co-Design Improvisation}

\author{
\IEEEauthorblockN{Alex Binh Vinh Duc Nguyen\orcidlink{0000-0001-5026-474X}}
\IEEEauthorblockA{
\textit{KU Leuven}\\
Leuven, Belgium \\
\href{mailto:alex.nguyen@kuleuven.be}{\nolinkurl{alex.nguyen@kuleuven.be}}}
\and
\IEEEauthorblockN{Jan Leusmann\orcidlink{0000-0001-9700-5868}}
\IEEEauthorblockA{
\textit{LMU Munich}\\
Munich, Germany \\
\href{mailto:jan.leusmann@ifi.lmu.de}{\nolinkurl{jan.leusmann@ifi.lmu.de}}}
\and
\IEEEauthorblockN{Sven Mayer\orcidlink{0000-0001-5462-8782}}
\IEEEauthorblockA{
\textit{LMU Munich}\\
Munich, Germany \\
\href{mailto:info@sven-mayer.com}{\nolinkurl{info@sven-mayer.com}}}
\and
\IEEEauthorblockN{Andrew Vande Moere\orcidlink{0000-0002-0085-4941}}
\IEEEauthorblockA{
\textit{KU Leuven}\\
Leuven, Belgium \\
\href{mailto:andrew.vandemoere@kuleuven.be}{\nolinkurl{andrew.vandemoere@kuleuven.be}}}
}

\maketitle


\vfill

\setcounter{table}{0}
\begin{table}[h]
    \centering
    \caption{Participant demographics as self-reported}
    \label{tab:demographics}
    \small
    \begin{tabularx}{0.9\linewidth}{clll}
    \toprule
    \textbf{Participant} & \textbf{Expertise} & \textbf{Age} & \textbf{Gender} \\
    \midrule
    P01 & Animation & 31 & Female \\
    P02 & Animation & 49 & Male \\
    P03 & Animation & 27 & Male \\
    \midrule
    P04 & Cinematography & 32 & Male \\
    P05 & Cinematography & 35 & Female \\
    \midrule
    P06 & Choreography & 24 & Male\\
    P07 & Choreography & 30 & Female \\
    \midrule
    P08 & Scenography & 50 & Male \\
    P09 & Scenography & 32 & Female \\
    \midrule
    P10 & Architectural design & 29 & Female \\
    P11 & Architectural design & 27 & Female \\
    P12 & Architectural design & 40 & Female \\
    \midrule
    P13 & Robotic development & 35 & Male \\
    P14 & Robotic development & 31 & Female \\
    P15 & Robotic development & 32 & Male \\
    \bottomrule
    \end{tabularx}
\end{table}

\setcounter{table}{1}
\begin{table}[t!]
    \centering
    \caption{Distribution of three gestural types among 15 participants}
    \label{tab:participantgestures}
    \small
    \begin{tabularx}{0.9\linewidth}{cccc}
    \toprule
    \textbf{Participant} & \textbf{Architectonic} & \textbf{Deictic} & \textbf{Emblematic} \\
    \midrule
    P01 & & 4 & 4 \\
    P02 & 1 & 3 & 3 \\
    P03 & 3 & & 3 \\
    \midrule
    P04 & 5 & & 1\\
    P05 & 1 & 1 & 4 \\
    \midrule
    P06 & & 4 & 4 \\
    P07 & 2 & 2 & 3 \\
    \midrule
    P08 & 2 & 1 & 3 \\
    P09 & 4 & 1 & 1 \\
    \midrule
    P10 & & 6 & 2 \\
    P11 & 3 & 2 & 3\\
    P12 & 5 & 1\\
    \midrule
    P13 & 2 & 3 & 2\\
    P14 & 1 & 2 & 3 \\
    P15 & 1 & 1 & 5 \\
    \bottomrule
    \end{tabularx}
\end{table}

\vfill

\newpage

\begin{figure*}[t!]
    \centering
    \includegraphics[width=\linewidth]{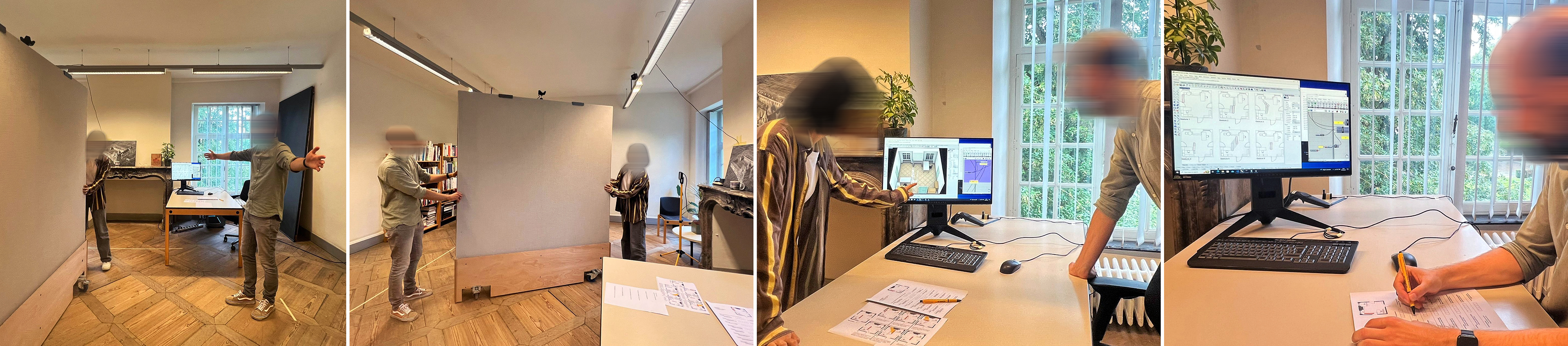}
    \caption{\textit{From left to right:} During improvisation sessions, 15 experts designed gestures for a mobile robotic partition by first bodystorming how to convey an intent (1), then manually manoeuvring the partition with a researcher to perform the accordant gesture (2), before reviewing the 3D visualisation of the gesture as recorded via a real-time motion-capture system (3), and comparing all designed gestures in a questionnaire while viewing their 2D animations side-by-side (4).
    }
    \label{fig:teaser_original}
\end{figure*}

\setcounter{table}{2}
\begin{table}[h]
    \centering
    \caption{Distribution of 20 gestural strategies\\among six intents}
    \label{tab:patternoccurence}
    \scriptsize
    \begin{tabularx}{0.9\linewidth}{lX|ll|ll|ll}
    \toprule
    \multicolumn{2}{c}{\textbf{Gestural strategy}} & \multicolumn{6}{c}{\textbf{Intent}} \\
    && I1 & I2 & I3 & I4 & I5 & I6 \\
    \midrule
    \multicolumn{8}{c}{\cellcolor{white!85!black}\textit{Architectonic strategies}} \\
    GA1 & \textbf{Directional corridor} &&& 2 & 2 \\
    GA2 & \textbf{Welcoming space} &&& 4 && \\
    GA3 & \textbf{Visual connection} &&&& 1 & & 1 \\
    \midrule
    GA4 & \textbf{Close impact} & 3 &&&&& \\
    GA5 & \textbf{Midway pause} & 2 &&&&& \\
    GA6 & \textbf{Slow reveal} & 3 &&&&& \\
    GA7 & \textbf{Block and open} &&&& 2 &&  \\
    \midrule
    \multicolumn{8}{c}{\cellcolor{white!85!black}\textit{Deictic strategies}} \\
    GD1 & \textbf{Diagonal direction} & \textbf{7} && \textbf{5} & \textbf{9} && \\
    \midrule
    GD2 & \textbf{Width back-off} && 2 &&&& \\
    GD3 & \textbf{Directional retreat} && 1 &&&& \\
    \midrule
    GD4 & \textbf{Length point}  &&& 3 &&&\\
    GD5 & \textbf{Hinged point} &&& 2 & 2 &&\\
    \midrule
    \multicolumn{8}{c}{\cellcolor{white!85!black}\textit{Emblematic strategies}} \\  
    GE1 & \textbf{Custom position} && 5 &&& 2 & 3 \\
    \midrule
    GE2 & \textbf{Confused back-and-forth} &&&&&& 2 \\
    GE3 & \textbf{Confused head shake} &&&&&& \textbf{6} \\
    GE4 & \textbf{Confused look around} &&&&&& 3 \\
    \midrule
    GE5 & \textbf{Length wave}  &&&& 5 & 2 \\
    GE6 & \textbf{Hinged wave} &&& 4 && \textbf{4} \\
    GE7 & \textbf{Symmetrical wave} && \textbf{7} && 1 & 3 \\
    GE8 & \textbf{Width step-up} &&&&& \textbf{4} \\
    \midrule
    \multicolumn{2}{r|}{\textit{Sum}} & 15 & 15 & 20 & 22 & 15 & 15 \\
    \bottomrule
    \end{tabularx}
\end{table}

\setcounter{table}{3}
\begin{table}[h]
    \centering
    \caption{Occurrence scores (\%) of 20 gestural strategies\\among six intents}
    \label{tab:patterndistribution}
    \scriptsize
    \begin{tabularx}{\linewidth}{lX|ll|ll|ll}
    \toprule
    \multicolumn{2}{c}{\textbf{Gestural strategies}} & \multicolumn{6}{c}{\textbf{Intent}} \\
    && I1 & I2 & I3 & I4 & I5 & I6 \\
    \midrule
    \multicolumn{8}{c}{\cellcolor{white!85!black}\textit{Architectonic strategies}} \\
    GA1 & \textbf{Directional corridor} &&& 10 & 9 \\
    GA2 & \textbf{Welcoming space} &&& 20 && \\
    GA3 & \textbf{Visual connection} &&&& 5 & & 7 \\
    \midrule
    GA4 & \textbf{Close impact} & 20 &&&&& \\
    GA5 & \textbf{Midway pause} & 13 &&&&& \\
    GA6 & \textbf{Slow reveal} & 20 &&&&& \\
    GA7 & \textbf{Block and open} &&&& 9 &&  \\
    \midrule
    \multicolumn{8}{c}{\cellcolor{white!85!black}\textit{Deictic strategies}} \\
    GD1 & \textbf{Diagonal direction} & \textbf{47} && \textbf{25} & \textbf{41} && \\
    \midrule
    GD2 & \textbf{Width back-off} && 13 &&&& \\
    GD3 & \textbf{Directional retreat} && 7 &&&& \\
    \midrule
    GD4 & \textbf{Length point}  &&& 15 &&&\\
    GD5 & \textbf{Hinged point} &&& 10 & 9 &&\\
    \midrule
    \multicolumn{8}{c}{\cellcolor{white!85!black}\textit{Emblematic strategies}} \\  
    GE1 & \textbf{Custom position} && 33 &&& 13 & 20 \\
    \midrule
    GE2 & \textbf{Confused back-and-forth} &&&&&& 13 \\
    GE3 & \textbf{Confused head shake} &&&&&& \textbf{40} \\
    GE4 & \textbf{Confused look around} &&&&&& 20 \\
    \midrule
    GE5 & \textbf{Length wave}  &&&& 23 & 13 \\
    GE6 & \textbf{Hinged wave} &&& 20 && \textbf{27} \\
    GE7 & \textbf{Symmetrical wave} && \textbf{47} && 5 & 20 \\
    GE8 & \textbf{Width step-up} &&&&& \textbf{27} \\
    \midrule
    \multicolumn{2}{r|}{\textit{Sum}} & 100 & 100 & 100 & 100 & 100 & 100 \\
    \bottomrule
    \end{tabularx}
\end{table}

\newpage

\begin{figure*}[h]
    \centering
    \includegraphics[width=\linewidth]{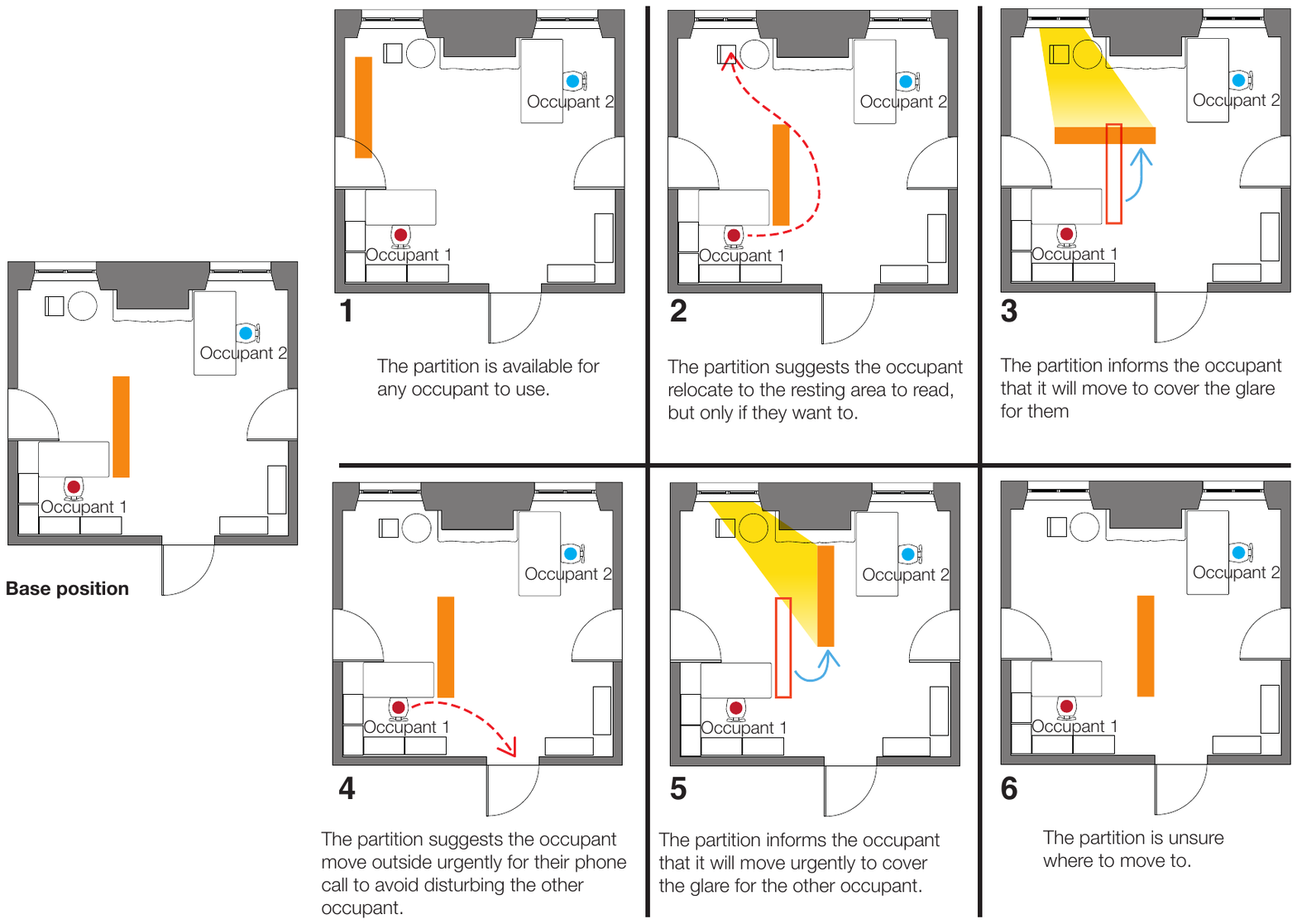}
    \caption{The printed sheet of the intent referents that were given to participants during improvisation sessions. As shown in this figure, the order of the intent referents were randomised for each participant.
    }
    \label{fig:catalogue}
\end{figure*}



\begin{figure*}[h]
    \centering
    \includegraphics[width=\linewidth]{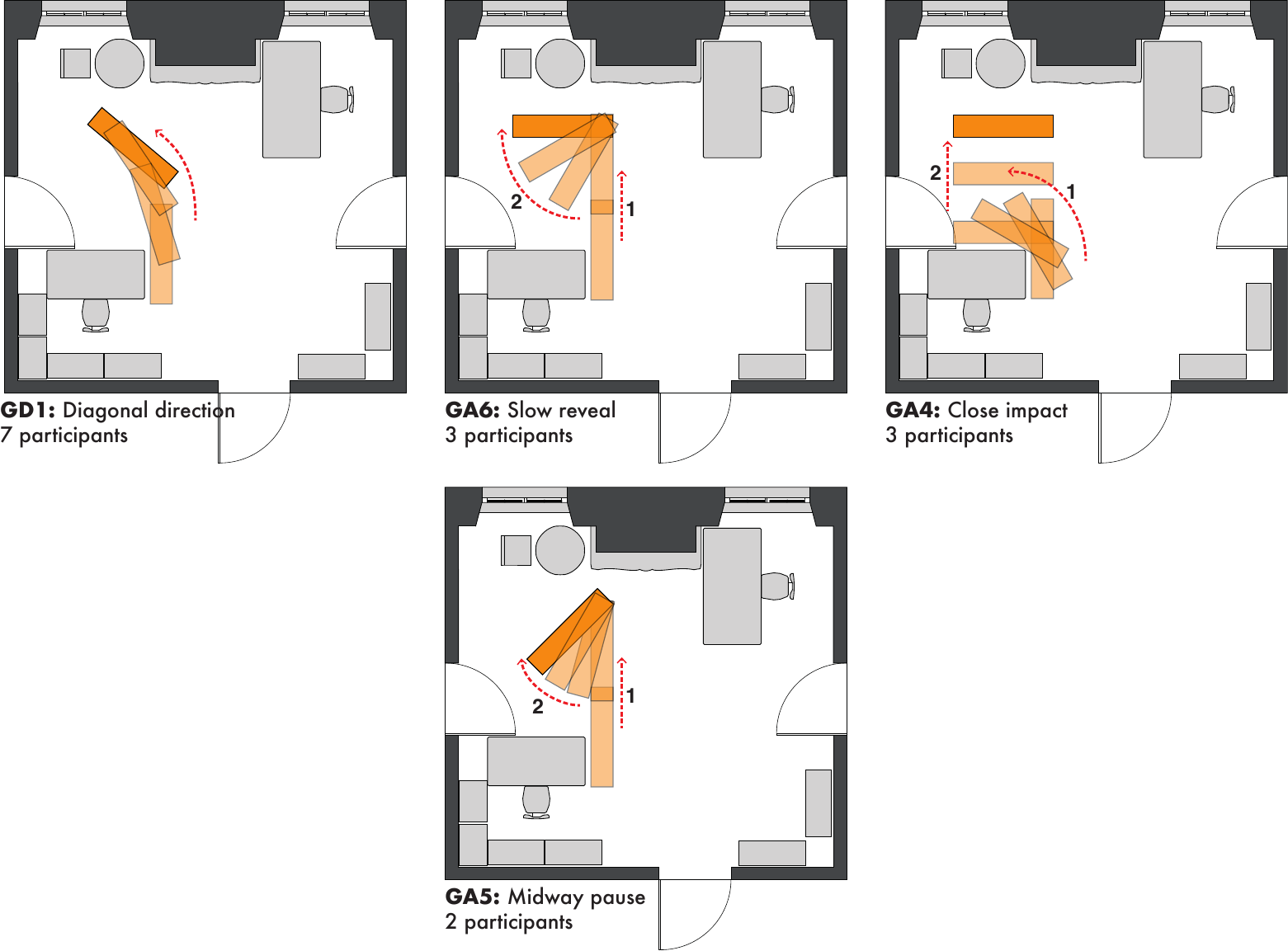}
    \caption{The gestural strategies that participants designed for \textbf{I1 (Informing / Non-urgent)}: ``The partition informs the occupant that it will move to cover the glare for them".
    }
    \label{fig:I1}
\end{figure*}

\begin{figure*}[h]
    \centering
    \includegraphics[width=\linewidth]{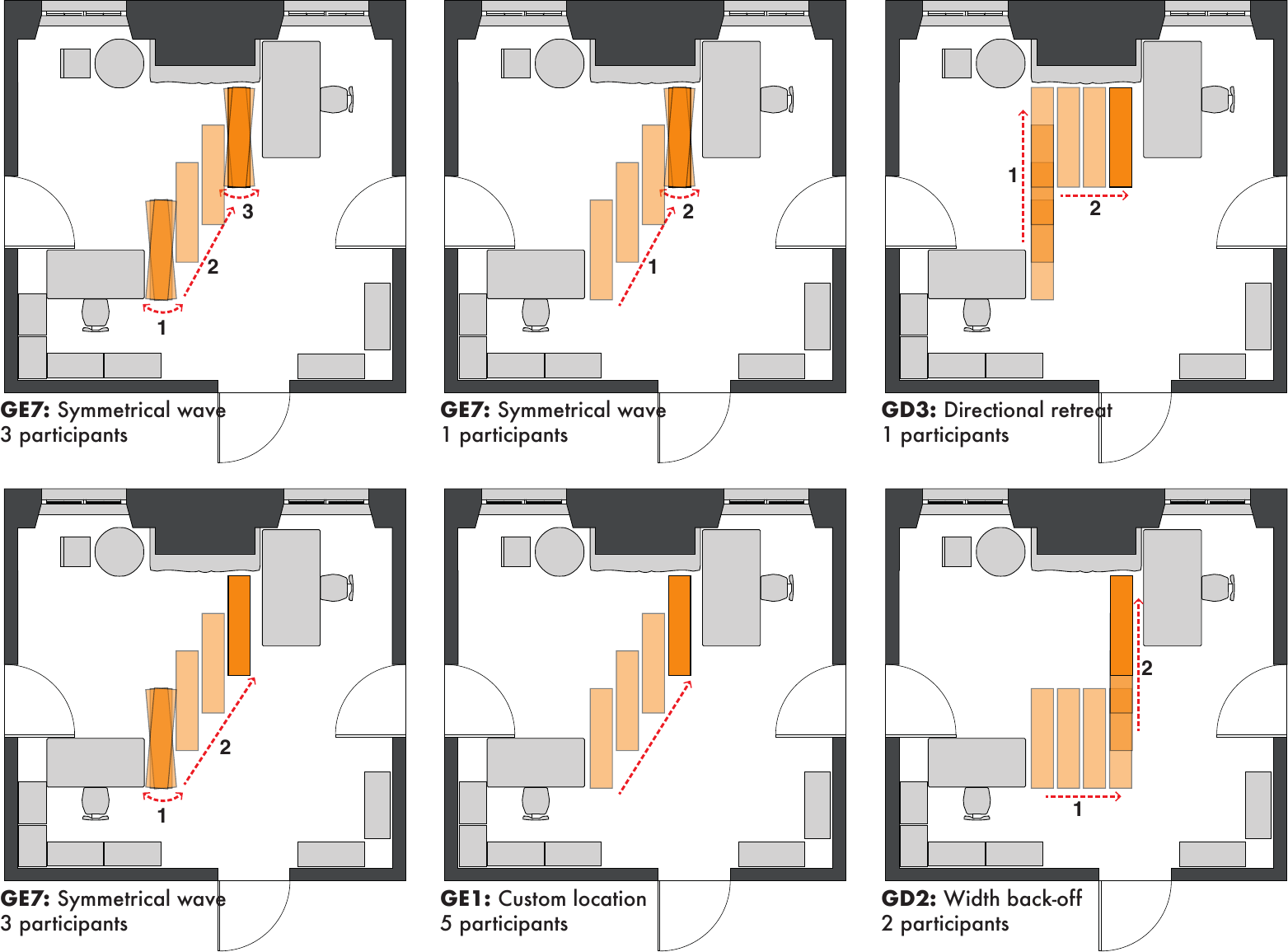}
    \caption{The gestural strategies that participants designed for \textbf{I2 (Informing / Urgent)}: ``The partition informs the occupant that it will move urgently to cover the glare for the other occupant".
    }
    \label{fig:I2}
\end{figure*}

\begin{figure*}[h]
    \centering
    \includegraphics[width=1\linewidth]{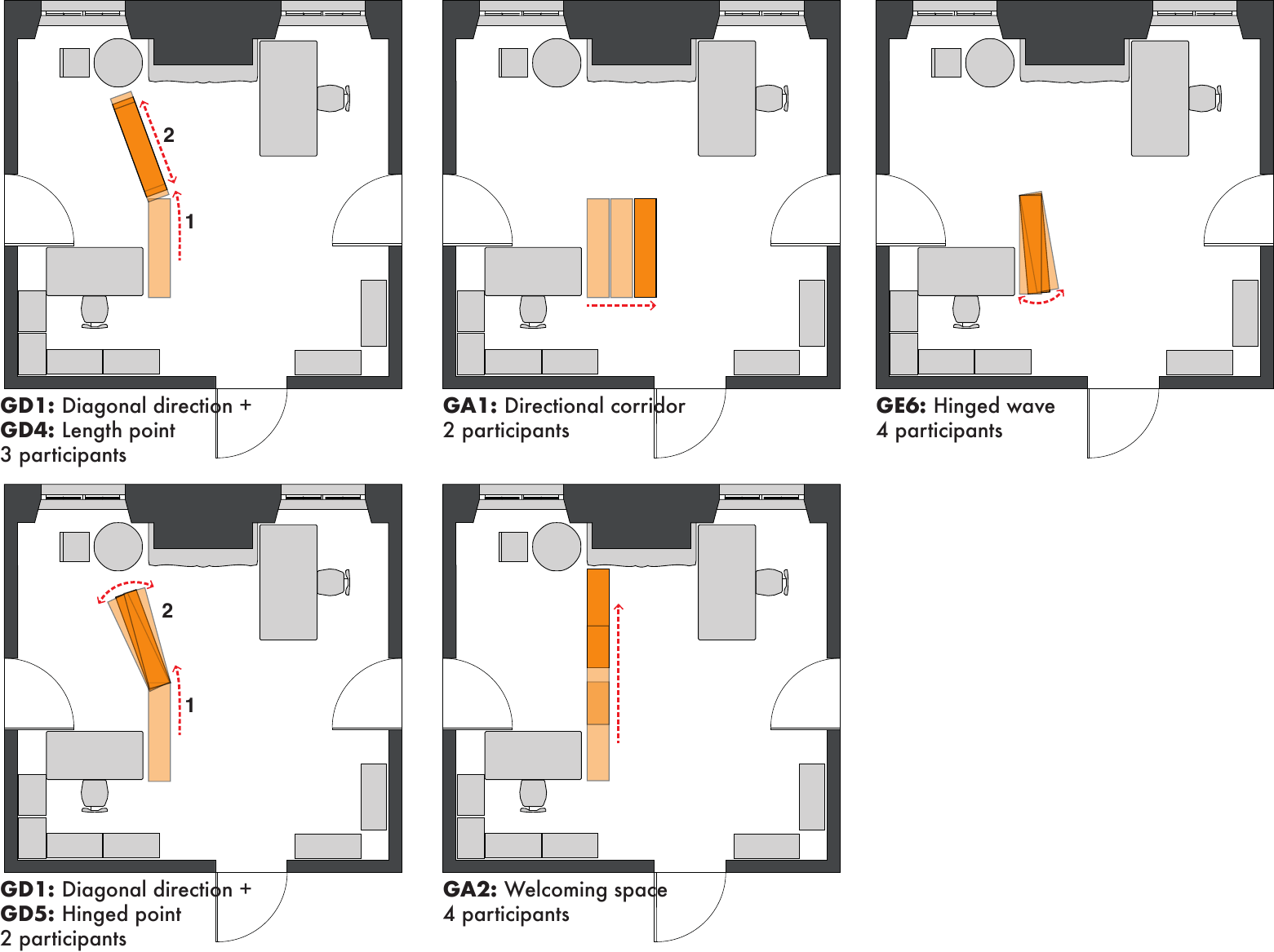}
    \caption{The gestural strategies that participants designed for \textbf{I3 (Nudging / Non-urgent)}: ``The partition suggests the occupant relocate to the resting area to read, but only if they want to.".
    }
    \label{fig:I3}
\end{figure*}

\begin{figure*}[h]
    \centering
    \includegraphics[width=1\linewidth]{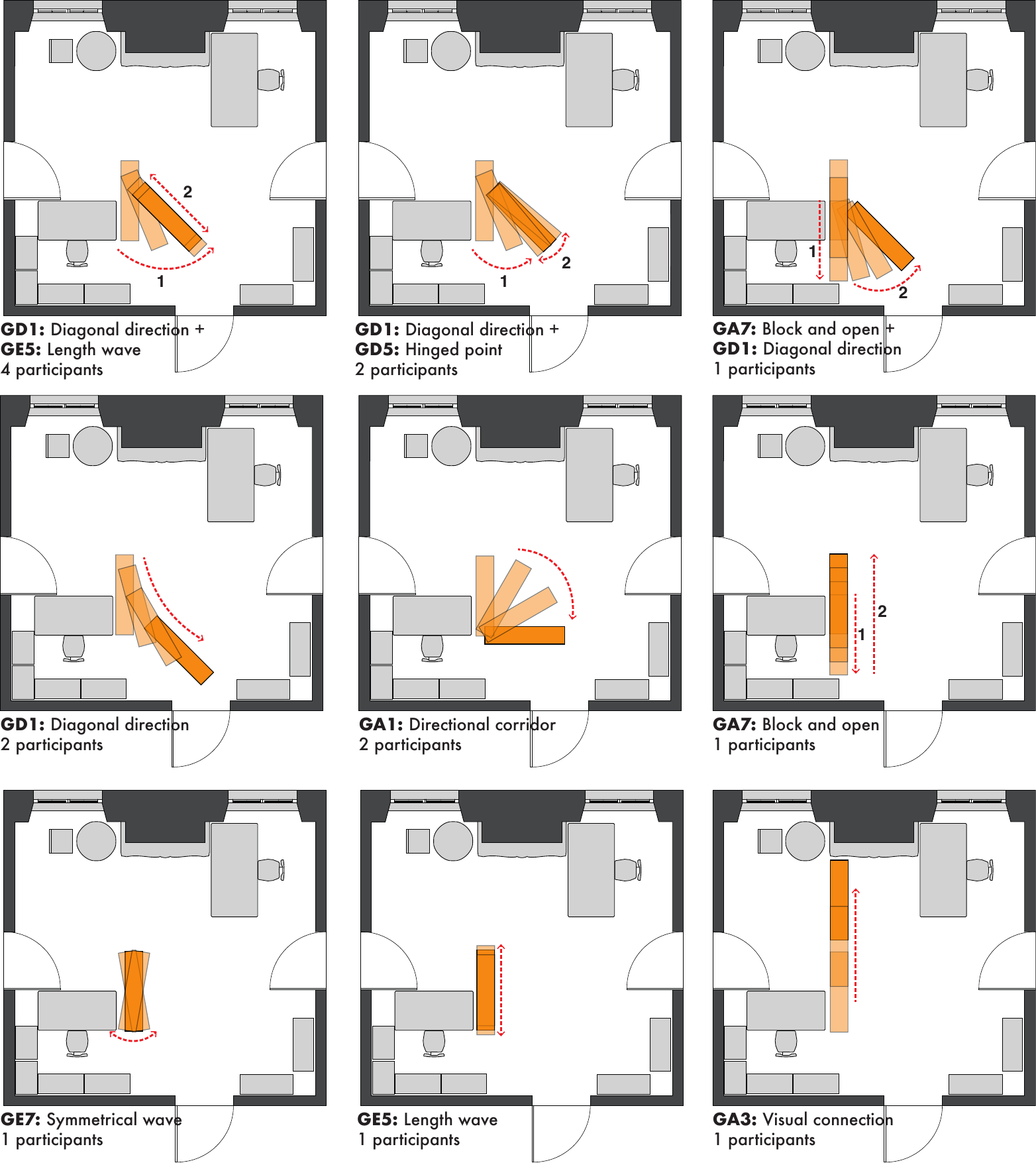}
    \caption{The gestural strategies that participants designed for \textbf{I4 (Nudging / Urgent)}: ``The partition suggests the occupant move outside urgently for their phone call to avoid disturbing the other occupant.".
    }
    \label{fig:I4}
\end{figure*}

\begin{figure*}[h]
    \centering
    \includegraphics[width=\linewidth]{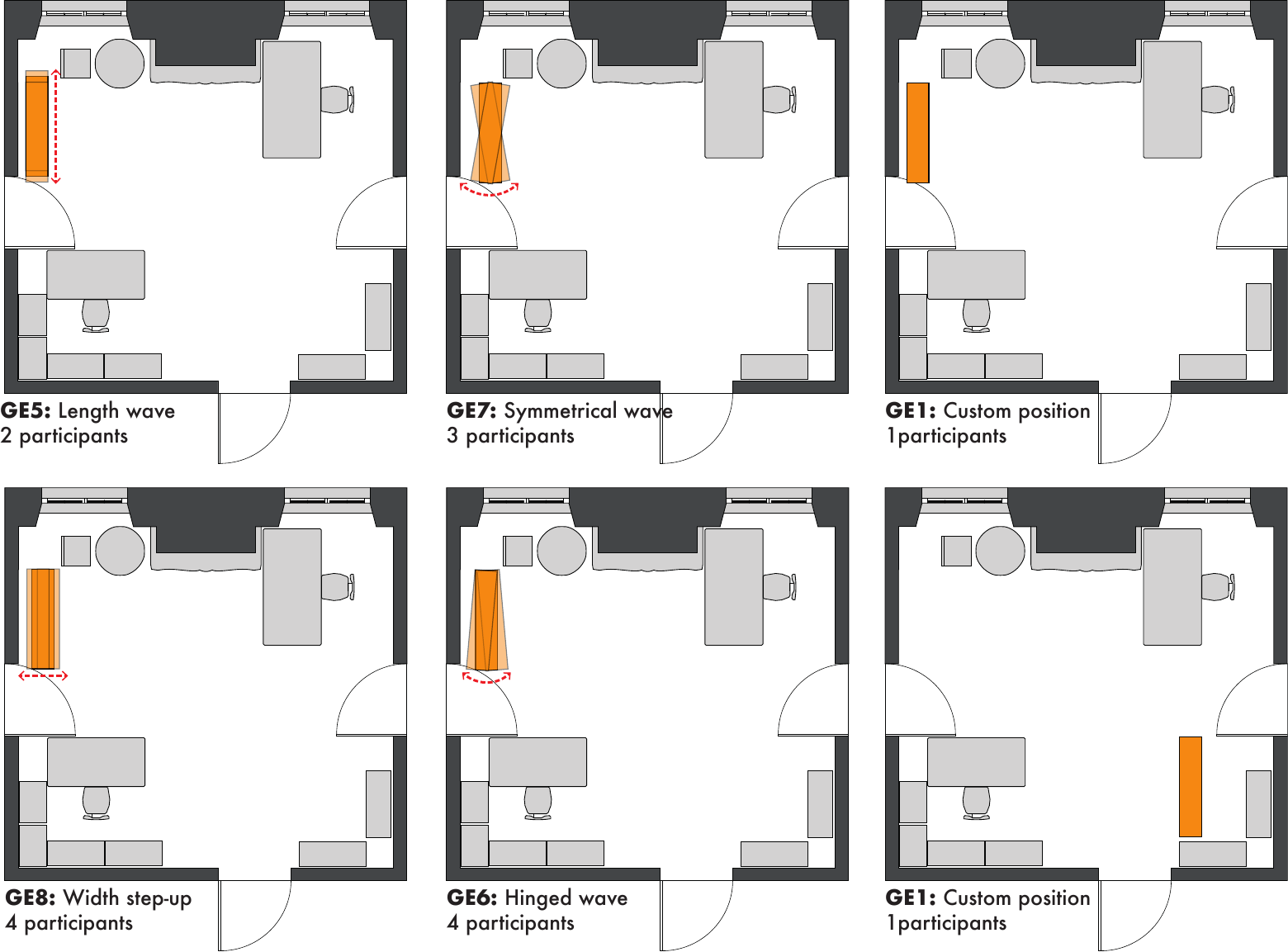}
    \caption{The gestural strategies that participants designed for \textbf{I5 (Control)}: ``The partition is available for any occupant to use".
    }
    \label{fig:I5}
\end{figure*}

\begin{figure*}[h]
    \centering
    \includegraphics[width=1\linewidth]{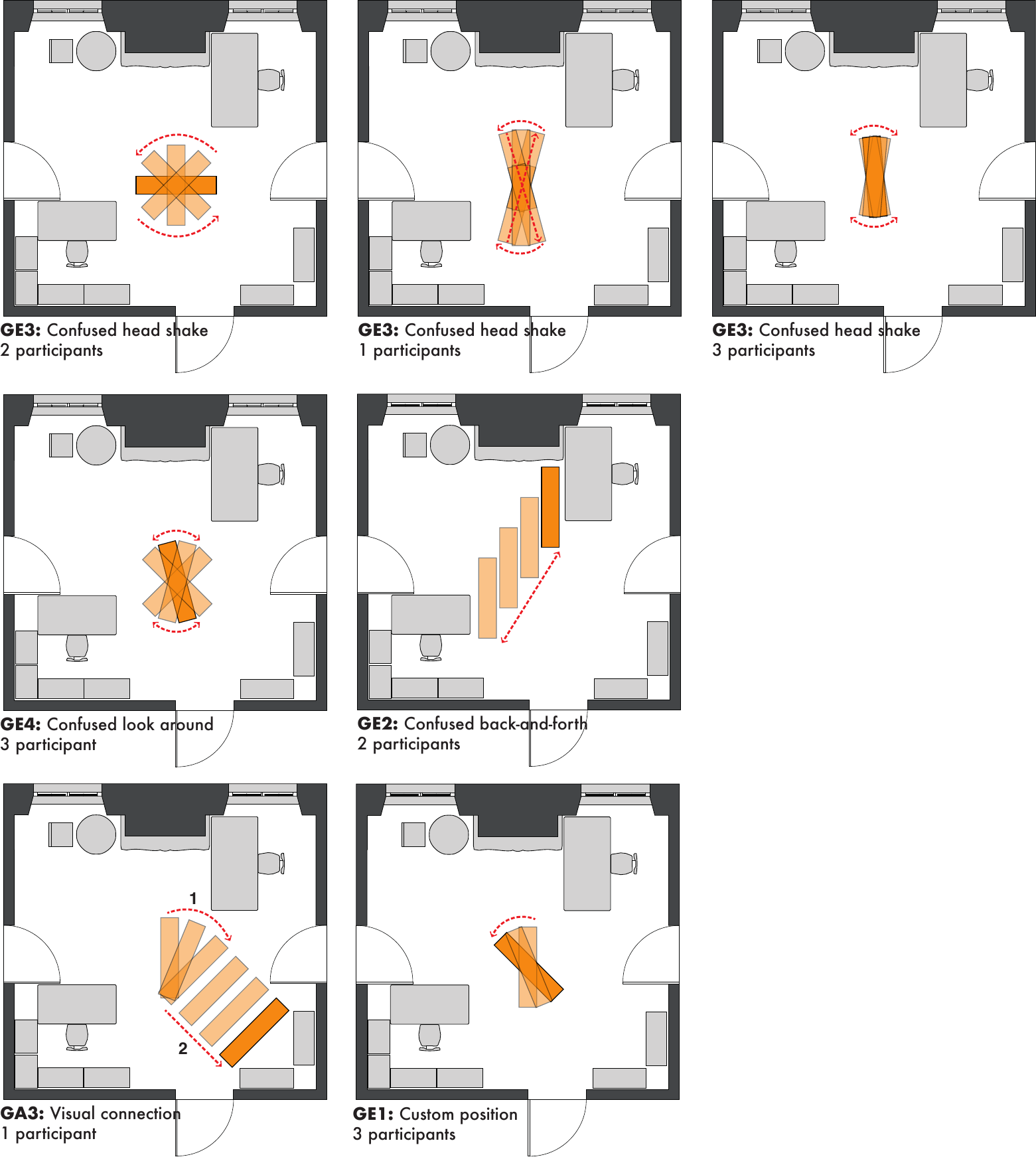}
    \caption{The gestural strategies that participants designed for \textbf{I6 (Control)}: ``The partition is unsure where to move to".
    }
    \label{fig:I6}
\end{figure*}


\begin{figure*}[h]
    \centering
    \includegraphics[width=0.8\linewidth]{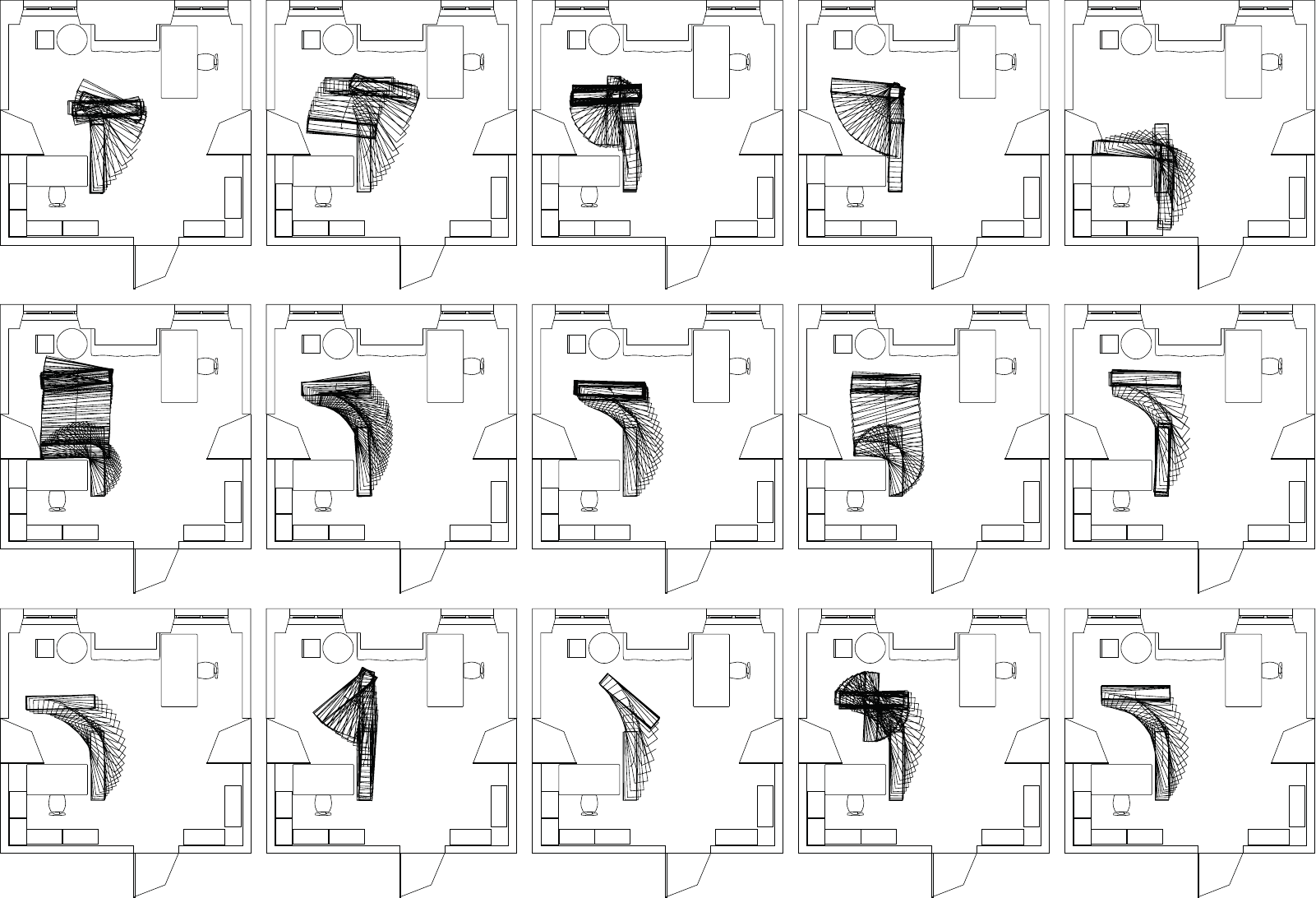}
    \caption{The designed gestures as recorded by the motion tracking system for \textbf{I1 (Informing / Non-urgent)}: ``The partition informs the occupant that it will move to cover the glare for them".
    }
    \label{fig:dataI1}
\end{figure*}

\begin{figure*}[h]
    \centering
    \includegraphics[width=0.8\linewidth]{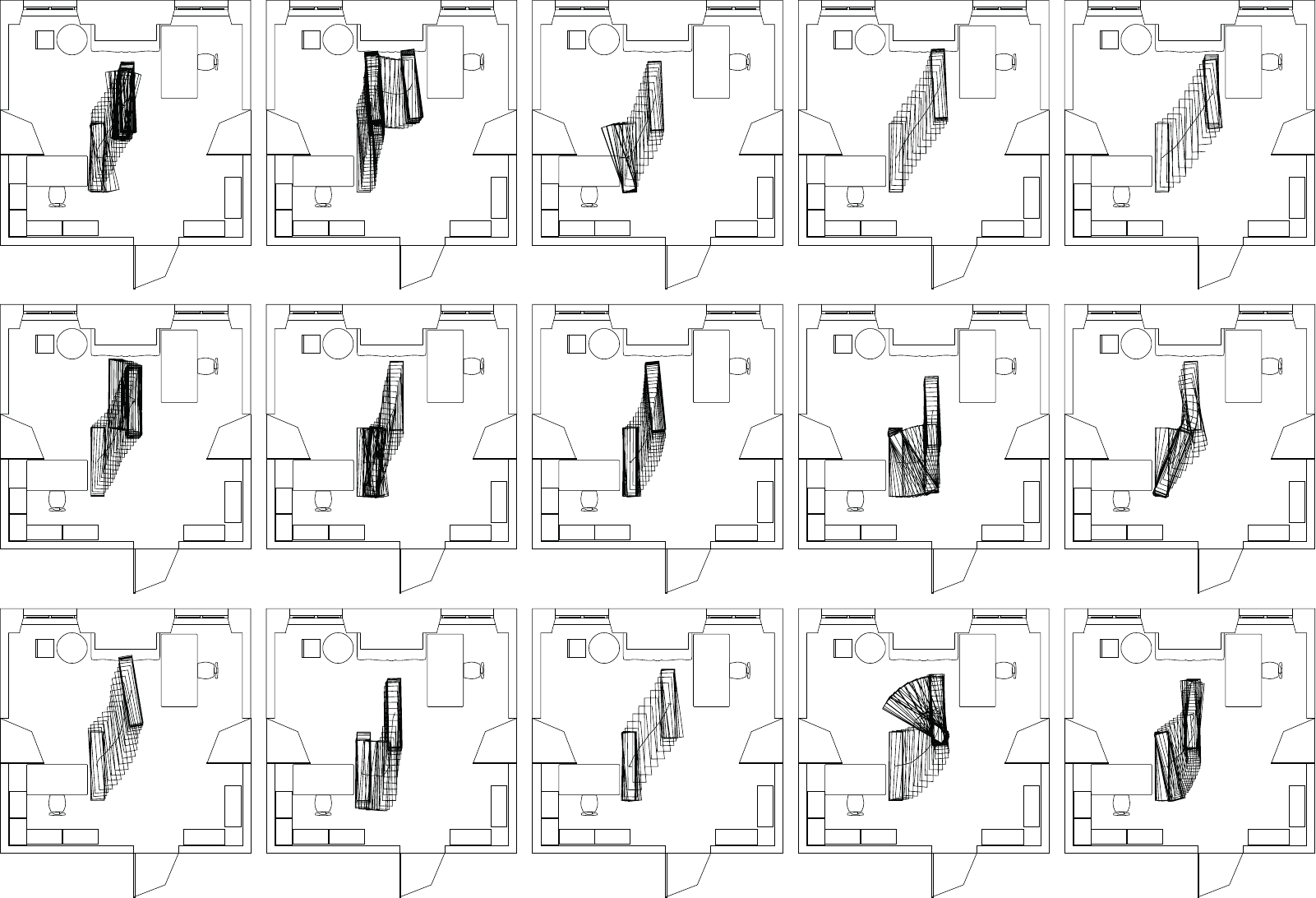}
    \caption{The designed gestures as recorded by the motion tracking system for \textbf{I2 (Informing / Urgent)}: ``The partition informs the occupant that it will move urgently to cover the glare for the other occupant".
    }
    \label{fig:dataI2}
\end{figure*}

\begin{figure*}[h]
    \centering
    \includegraphics[width=0.8\linewidth]{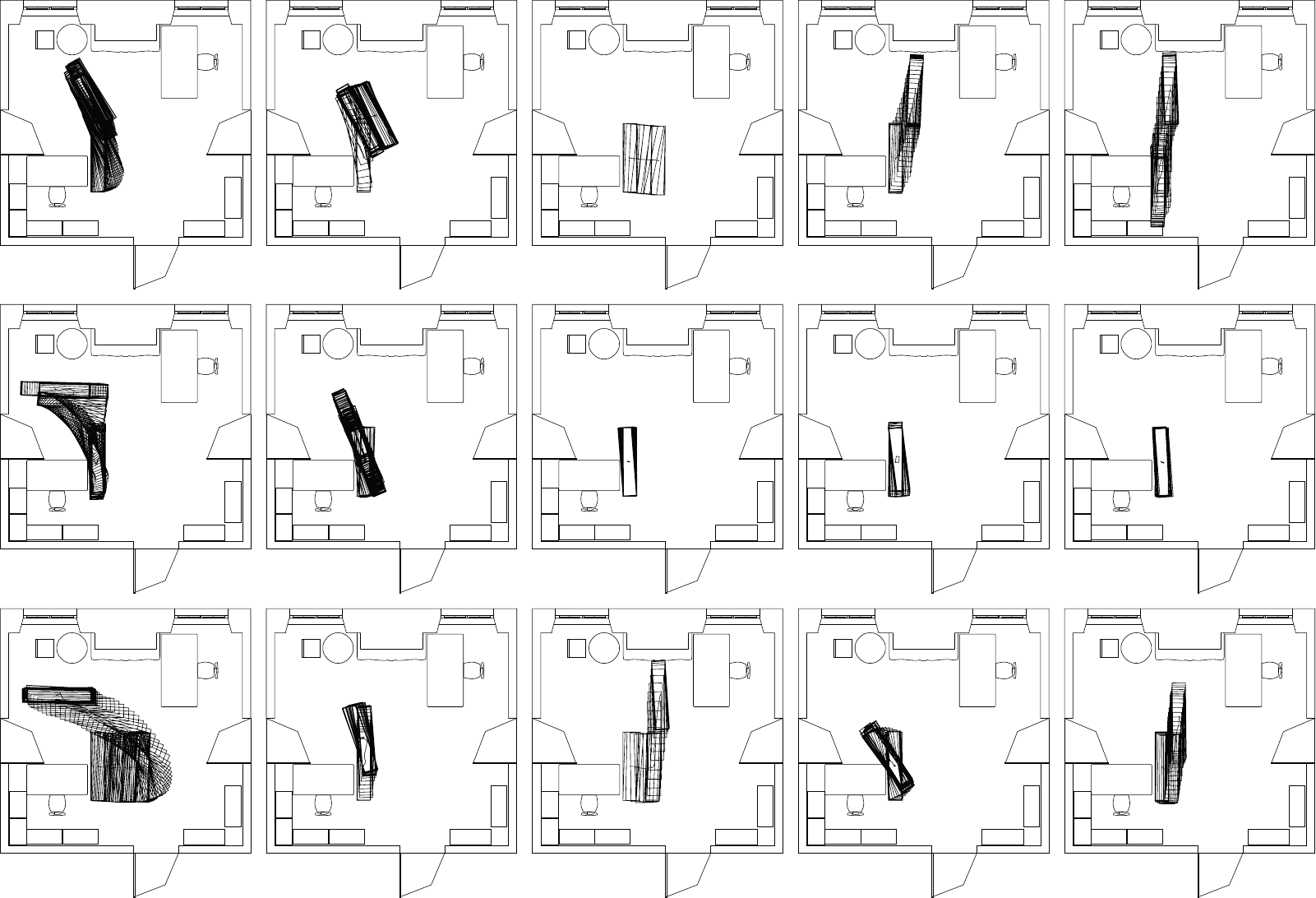}
    \caption{The designed gestures as recorded by the motion tracking system for \textbf{I3 (Nudging / Non-urgent)}: ``The partition suggests the occupant relocate to the resting area to read, but only if they want to.".
    }
    \label{fig:dataI3}
\end{figure*}

\begin{figure*}[h]
    \centering
    \includegraphics[width=0.8\linewidth]{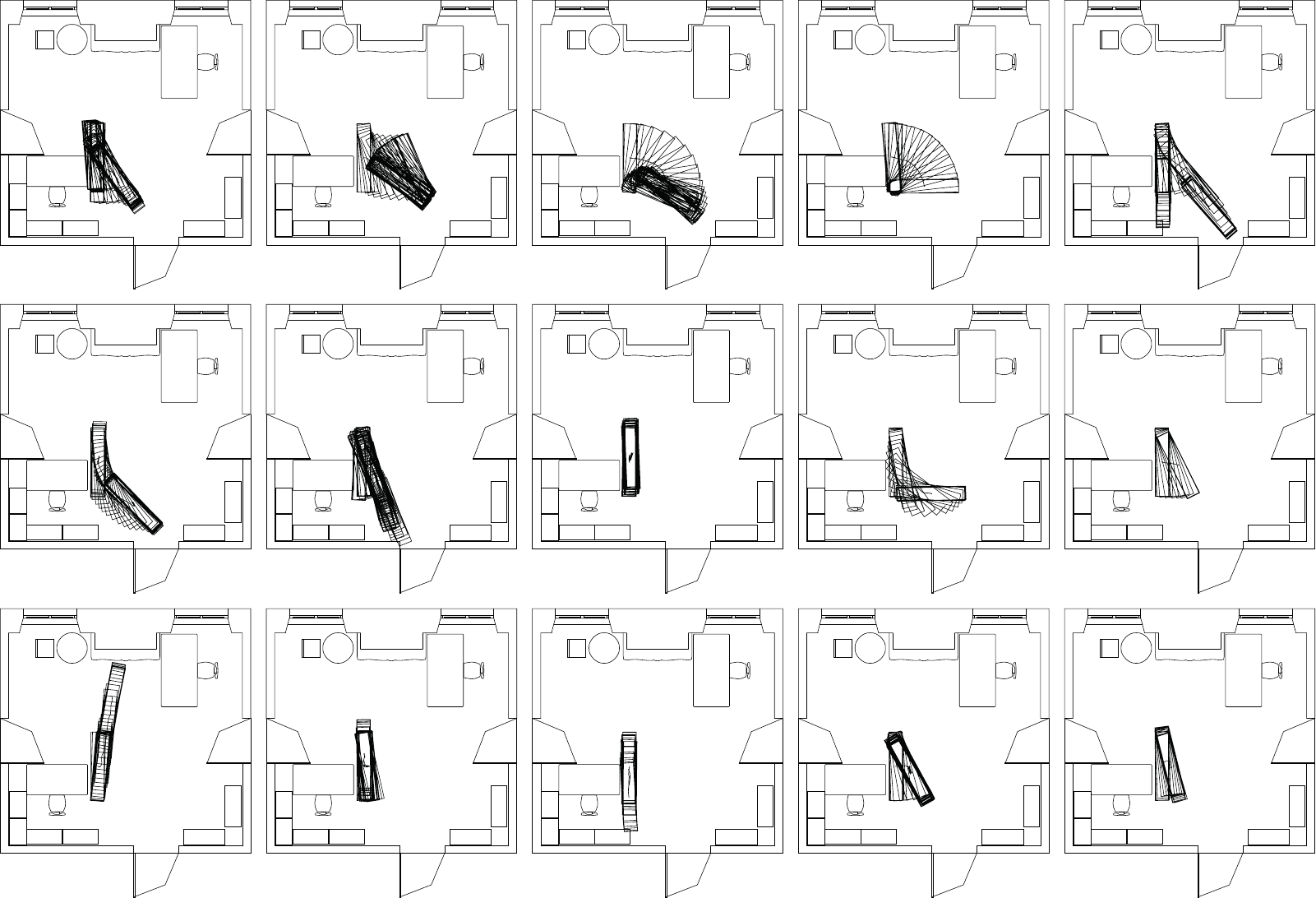}
    \caption{The designed gestures as recorded by the motion tracking system for \textbf{I4 (Nudging / Urgent)}: ``The partition suggests the occupant move outside urgently for their phone call to avoid disturbing the other occupant.".
    }
    \label{fig:dataI4}
\end{figure*}

\begin{figure*}[h]
    \centering
    \includegraphics[width=0.8\linewidth]{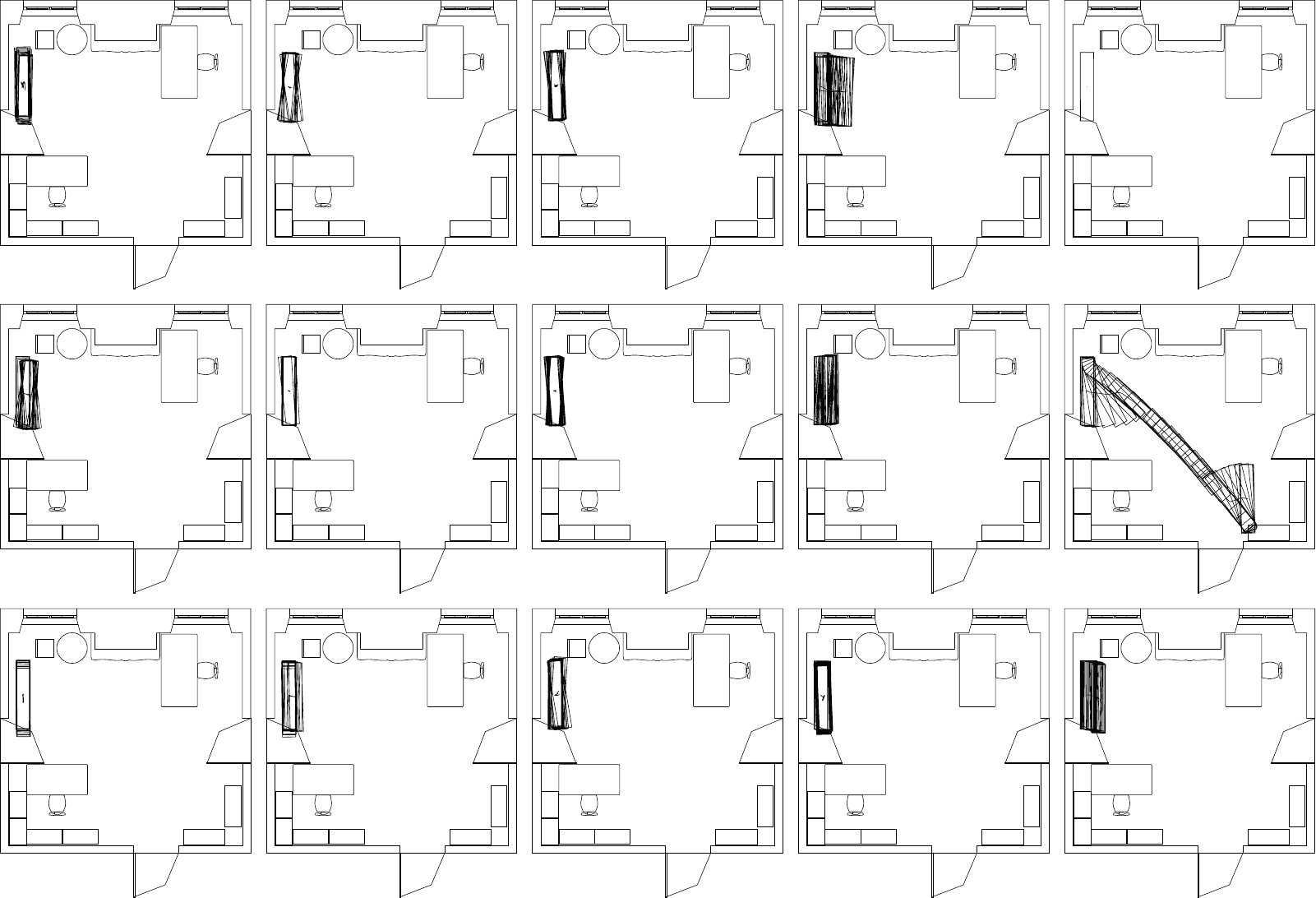}
    \caption{The designed gestures as recorded by the motion tracking system for \textbf{I5 (Control)}: ``The partition is available for any occupant to use".
    }
    \label{fig:dataI5}
\end{figure*}

\begin{figure*}[h]
    \centering
    \includegraphics[width=0.8\linewidth]{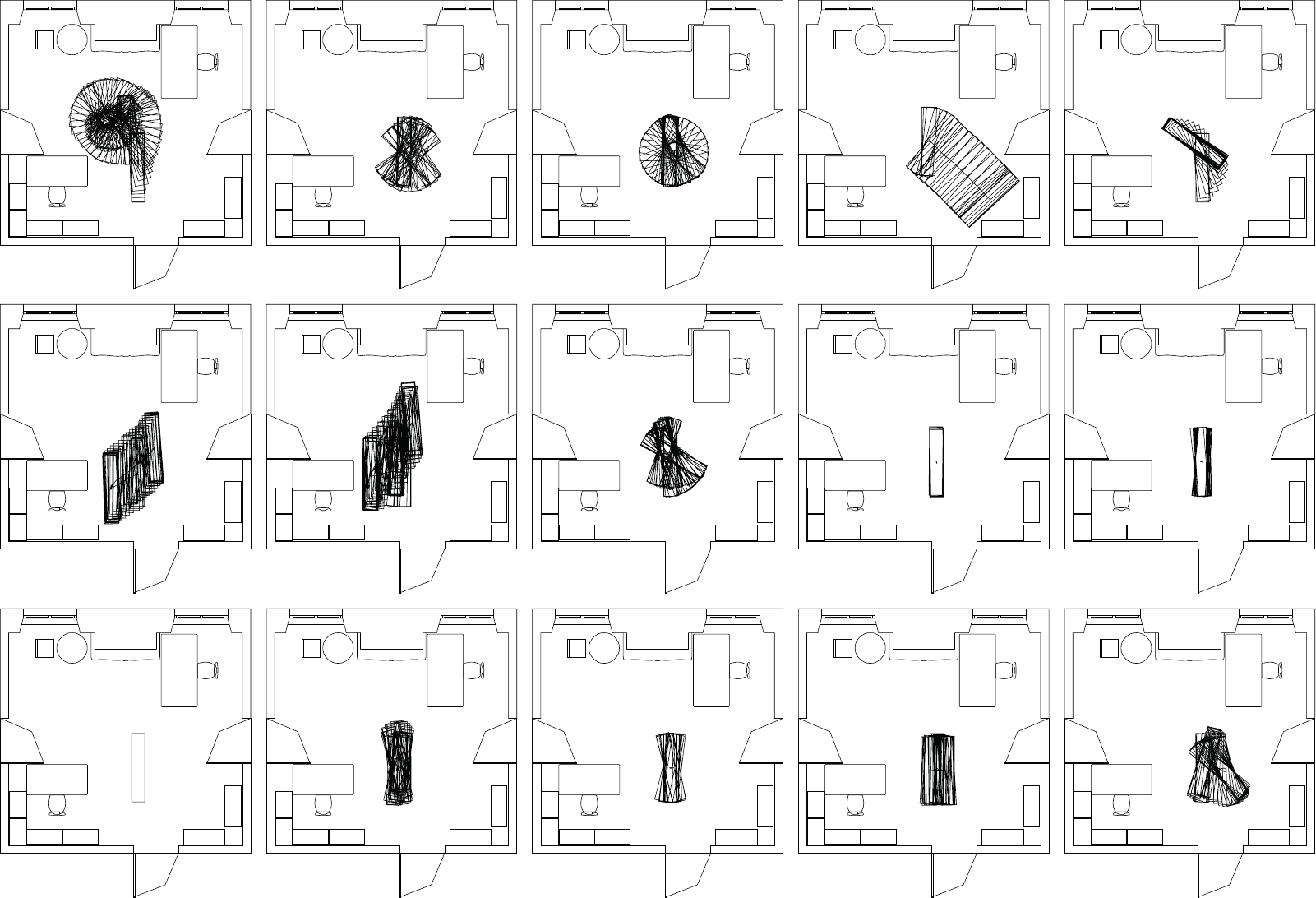}
    \caption{The designed gestures as recorded by the motion tracking system for \textbf{I6 (Control)}: ``The partition is unsure where to move to".
    }
    \label{fig:dataI6}
\end{figure*}

